\documentclass[twocolumn,aps,prb,final,amsfonts,
amssymb,amsmath,floatfix,showpacs,superscriptaddress]{revtex4}

\usepackage{graphicx}
\usepackage{bm}
\begin{document}

\title{The effects of macroscopic inhomogeneities on the magneto
transport properties of the electron gas in two dimensions}
\author{B.~Karmakar}
\email{karmakar@tifr.res.in}
\affiliation{Tata Institute of Fundamental Research, Mumbai 400005, India.}
\author{M.~R.~Gokhale}
\affiliation{Tata Institute of Fundamental Research, Mumbai 400005, India.}
\author{A.~P.~Shah}
\affiliation{Tata Institute of Fundamental Research, Mumbai 400005, India.}
\author{B.~M.~Arora}
\affiliation{Tata Institute of Fundamental Research, Mumbai 400005, India.}
\author{D.~T.~N.~de~Lang}
\affiliation{Van der Waals-Zeeman Institute, University of Amsterdam, The
Netherlands}
\author{A.~de~Visser}
\affiliation{Van der Waals-Zeeman Institute, University of Amsterdam, The
Netherlands}
\author{L.~A.~Ponomarenko}
\affiliation{Van der Waals-Zeeman Institute, University of Amsterdam, The
Netherlands}
\author{A.~M.~M.~Pruisken}
\email{pruisken@science.uva.nl}
\affiliation{Institute of Theoretical Physics, University of
Amsterdam, The Netherlands.}
\date{\today}

\begin{abstract}
In experiments on electron transport the {\em macroscopic
inhomogeneities} in the sample play a fundamental role. In this
paper and a subsequent one we introduce and develop a general
formalism that captures the principal features of sample
inhomogeneities ({\em density gradients, contact misalignments})
in the magneto resistance data taken from low mobility
heterostructures. We present detailed assessments and experimental
investigations of the different regimes of physical interest,
notably the regime of {\em semiclassical transport} at weak
magnetic fields, the {\em plateau-plateau} transitions as well as
the {\em plateau-insulator} transition that generally occurs at
much stronger values of the external field only.

It is shown that the semiclassical regime at weak fields plays an
integral role in the general understanding of the experiments on
the quantum Hall regime. The results of this paper clearly
indicate that the plateau-plateau transitions, unlike the the
plateau-insulator transition, are fundamentally affected by the
presence of sample inhomogeneities. We propose a {\em universal
scaling result} for the magneto resistance parameters. This result
facilitates, amongst many other things, a detailed understanding
of the difficulties associated with the experimental methodology
of H.P. Wei {\em et.al} in extracting the {\em quantum critical}
behavior of the electron gas from the transport measurements
conducted on the plateau-plateau transitions.

\end{abstract}


\maketitle


\section{Introduction}

Quantum phase transitions in the quantum Hall regime were first
studied by H.P. Wei {et al.}~\cite{wei:prl88} who demonstrated
that the {\em width} $\Delta\nu (T)$ of the transitions between
adjacent quantum Hall plateaus (PP transitions) becomes infinitely
narrow as the temperature $T$ goes to zero. An algebraic behavior
$\Delta\nu (T) \propto T^\kappa$ with $\kappa = 0.42 \pm 0.04$ was
found, independent of the Landau level\cite{wei:prl88,hwang:prb93}.
This indicated that the PP transitions in the quantum Hall regime
should be regarded as a {\em quantum critical phenomenon} with
universal critical indices\cite{pruisken:prl88}.

An overwhelming majority of subsequent experiments, by many
others\cite{koch,shahar:ssc97,hilke:nat98,pi_expts}, have
primarily revealed the elusive character of the quantum phase
transition. These experiments, unlike those by H.P. Wei {\em et.
al.}, were primarily conducted on arbitrarily chosen samples that
did not provide any access to the quantum critical phenomenon,
given the well known experimental limitations in reaching $T=0$.
It is well understood, although still not generally recognized as
such, that in order to be able to experimentally probe the true
$T=0$ asymptotics of the quantum phase transition, the dominant
scattering mechanism of the electron gas should be provided by
{\em short ranged potential fluctuations}. Smoothly varying
potential fluctuations (relative to the magnetic length) like
those present in the $GaAs/AlGaAs$ heterostructure mainly give
rise to an anomalously large {\em crossover} length scale for
scaling such that quantum criticality cannot be retrieved in the
experiment which is conducted at finite $T$. Crossover phenomena,
especially those between {\em percolation} and {\em localization},
have in general remained difficult to grasp. Nevertheless, they do
exist as an integral chapter in the renormalization theory of the
quantum Hall effect\cite{paperIII}. These phenomena play a crucial
and fundamental role in the understanding of not only laboratory
experiments but also the numerical experiments on the plateau
transitions in the quantum Hall regime.

Yet another complication is well known to play a major role, the
presence of macroscopic sample inhomogeneities that are inherent
to the experiments. Notice that spatial variations in the electron
density mainly produce a spatially varying filling fraction $\nu$
of the Landau level. Any such macroscopic inhomogeneity in the
electron density, no matter how small, will eventually complicate
the critical behavior of the electron gas in the limit where $T$
and, hence, the {\em width} $\Delta\nu(T)$ of the plateau
transitions goes to zero. The experimental situation is in many
ways similar to that of an ordinary liquid-gas phase transition
where, as is well known, inhomogeneity effects due to gravity
prevents one from entering arbitrary deeply into the critical
phase. Unlike the liquid gas phase transition, however, there
hardly exists any detailed study or systematic knowledge on the
inhomogeneity problem, especially in low mobility
heterostructures. Transport measurements on the Hall bar geometry
at low $T$ usually give rise to rather different results depending
on the pairs of contacts that are being used, the polarity of the
external field $B$ etc. Besides these geometrical aspects one
generally observes also slight differences in the data taken from
different experimental runs, before and after the sample has been
heated up to room temperature ($T$) and then cooled down
again\cite{schaijk:prl00}.

These annoying and puzzling complications have been the primary
reason why the experiments on the Hall bar geometry have so far
not provided any reliable information on the details of the {\em
scaling functions} of the conductivity parameters in the
transition regime between adjacent quantum Hall plateaus, notably
the peak value and the shape of $\sigma_{xx}$. Moreover, the most
important and fundamental aspect of the problem, the numerical
value of the critical index $\kappa$, has remained an unsettled
experimental problem. In spite of the fact that the original data
of H.P. Wei {\em et.al.} have provided an impressive experimental
demonstration of a quantum phase transition in the quantum Hall
regime, over the largest possible range of experimental $T$, it
has remained somewhat uncertain whether the extracted value of the
exponent $\kappa = 0.42 \pm 0.04$ is in fact the true {\em
critical} value, or whether it represents an {\em effective}
exponent resulting from an admixture of quantum critical behavior
and sample dependent effects due to macroscopic inhomogeneities.

Unexpected insights into the problem have been obtained more
recently\cite{pruisken:stretch}, as a result of a series of
detailed studies on quantum transport taken from the lowest Landau
level (PI transition) of a low mobility $InP-InGaAs$
heterostructure\cite{schaijk:prl00,lang:phe02}, Quite contrary to
the general statement which says that the basic phenomenon of
quantum criticality is independent of the Landau level, there are,
from an experimental point of view, fundamental and surprising
differences. It turns out that the {\em plateau-insulator} or PI
transition is generally much less affected by sample
inhomogeneities than the {\em plateau-plateau} or PP transitions
taken from the same sample. The difference primarily reveals
itself in terms of previously unrecognized {\em symmetries} in the
observed magneto resistance data under a change in the {\em
polarity} of the external magnetic field $B$. Moreover, the PI
transition, unlike the PP transitions, displays very specific and
general features that permit one to disentangle the quantum
critical aspects of the transport data (particle-hole symmetry,
critical exponents, scaling functions) and those that are
non-universal and sample dependent (gradients in the electron
density, contact misalignments). As a result of all this, one can
now say that the previously accepted experimental value of the
exponent $\kappa = 0.42$ is slightly incorrect. Instead, a new
value has been obtained, $\kappa = 0.57$. This new result
indicates that the quantum critical phenomenon belongs to a new,
non-Fermi-liquid universality class, in agreement with the more
recent advancements of the theory of localization and interaction
effects in the quantum Hall regime\cite{EuroLett,Finkel,twoloop}. At the
same time, the new studies also indicate that the {\em scaling
functions} of the transport parameters $\sigma_{xx}$ and
$\sigma_{xy}$ are, in fact, universal and in accordance with the
statement of self-duality under the Chern Simons mapping.

These surprising advances illuminate the problem of quantum
criticality in the quantum Hall regime by revealing all the
features sought in the PP transition that previously remained
concealed. In this paper we elaborate on the new insights that one
has gained from the studies of the PI transition and embark on the
specific difficulties associated with the PP transitions. We set
up a general, conceptual framework that enables one to not only
recognize the effects of macroscopic inhomogeneities from the
experimental data, but also perform the appropriate quantitative
analyses. Following the findings of Ref.\onlinecite{pruisken:stretch}
there are two distinctly different physical mechanisms at work in
the experiments. These mechanisms are

\begin{enumerate}
\item
A linear gradient in the electron density across the Hall bar,
\item
A misalignment of the contacts on the Hall bar.
\end{enumerate}

Our main objective is to demonstrate, both experimentally and
theoretically, that these physical mechanisms have a quite general
significance for the electron gas. They describe the inhomogeneity
effects in the transport data of not only the PI transition at
strong values of $B$, but also the PP transitions at
intermediately strong $B$ as well as the regime of semiclassical
transport (or weak quantum interference regime) that generally
occurs at weak values of $B$. We will show that the aforementioned
mechanisms manifest themselves quite differently, with distinctly
different physical consequences, in each regime of $B$ that one is
interested in.

As a general remark we can say that the experiments on the PP
transitions are most dramatically affected, even by relatively
weak inhomogeneities in the electron density. Whereas in practice
the complications are much less severe for both the PI transition
and the semiclassical regime and the corrections are relatively
simple, this is generally not the case for the PP transitions. The
main reason is that the effects of density gradients appear with a
strength proportional to $\partial \rho_{xy} / \partial B$, i.e.
the derivative of the Hall resistance with respect to the external
field. This quantity tends to diverge at the center of the PP
transition as $T$ approaches absolute zero. These effects,
however, are very weak in the semiclassical regime and even absent
near the PI transition where the Hall resistance $\rho_{xy}$
remains quantized and equal to $h / e^2$ at low $T$. This, then,
is the basic answer to the inhomogeneity problem stated at the
outset. It summarizes, at the same time, the principal conclusions
of this paper.

We already mentioned the fact that {\em symmetries} play an
important role in this magneto resistance problem. For example,
the recent analysis of the PI transition is based, to a large
extend, on an idea borrowed from the renormalization theory of the
quantum Hall effect which says that {\em particle-hole} symmetry
is a fundamental aspect of quantum criticality in the quantum Hall
regime\cite{pruisken:prl88}. This symmetry indicates, crudely
speaking, that the lines $\sigma_{xy} = half-integer$ should
emerge as axes of symmetry in the $\sigma_{xx}$ and $\sigma_{xy}$
conductivity plane. It has been known for a long time, however,
that this symmetry is generally violated in the experiment at low
$T$, due to the presence of macroscopic inhomogeneities in the
sample\cite{wei:prb86}.

As far as the PP transitions are concerned, the standard way of
dealing with this problem has been to "average" over the transport
data taken at opposite polarities of the external field $B$. This
procedure cannot always be taken seriously, however, because the
raw experimental data, especially at low $T$, usually deviates
substantially from the "averaged" value. In this paper we
elaborate on yet another, previously unrecognized symmetry of the
the PP transition\cite{pono} that enables one to distinguish
amongst the effects of {\em density gradients} and, say, {\em
contact misalignments}. This symmetry which we term {\em
reflection symmetry} indicates that no new information is added to
the Hall bar measurements if one changes the polarity of the $B$
field. More specifically, it says that under the change $B
\rightarrow -B$ the different resistances, taken from the four
terminals of the Hall bar, are in effect mapped onto one another.

We show that reflection symmetry is primarily the result of having
linear gradients in the electron density across the Hall bar and,
therefore, it generally stands for an approximate symmetry only.
Nevertheless, it has experimental significance even when
relatively large density gradients are present in the sample. Most
importantly, however, it teaches us how to proceed in order to
physically understand and describe the difficult inhomogeneity
problems that are associated with the PP transitions.

This paper is organized as follows. In Section 2 we outline a
theoretical analysis of density gradients in Hall bar samples on
the results of magneto transport experiments. Initially we
consider linear density gradients (Section 2.2) and obtain
explicit correction terms for the longitudinal and transverse
resistances. These reveal an important property - {\em reflection
symmetry} - in the magneto resistances taken from different
contact pairs, with reversal of the magnetic field polarity. These
results are directly applicable to the transport measurements on
the {\em semiclassical regime} (Section 3.2). Going beyond the
linear approximation, Section 2.3 presents the result of an
exactly solvable nonlinear problem involving exponential density
gradients. This analysis provides estimates of higher order
corrections due to nonlinear gradients in the semiclassical
regime. Gradient effects and related symmetry properties in the
high $B$ regime are also obtained. In Section 2.4 we explore the
sensitivity of the quantum critical behavior at the PP and PI
transitions to density gradients. This leads to specific limits on
the maximum density gradients which must be satisfied for
meaningful investigations of quantum criticality. Finally, in
Section 2.5, we discuss the recently reported experimental results
on universality of the scaling functions of the PI transitions. We
show, in particular, how they become generally useful in dealing
with the more difficult problems that are associated with the
experiments on the PP transitions. For this purpose we
recapitulate some of the principle advances in the theory of
localization and interaction effects in the quantum Hall regime
(Appendix) and propose explicit scaling results for the PP
transitions (Eqs 84 and 85).

In Section 3 we apply the results of Section 2 to the
experimental resistance data taken from a low mobility $InGaAs$
quantum well. The low B, semiclassical transport data display many
features expected from samples having linear density gradients,
most prominent being the {\em reflection symmetries}. We show how
the true transport parameters can be extracted that are free from
inhomogeneity effects. These in turn are shown to be useful, in
Section 3.2, for the investigation of (i) the departure from the
semicircle law in the $\sigma_{xx}$, $\sigma_{xy}$ conductivity
plane, (ii) the effects of weak quantum interference and (iii) the
effects of higher order, nonlinear behavior due to density
gradients. Section 3.3 describes the experimental results on the
quantum Hall regime, bringing out the symmetry properties, as well
as the limitations posed by the large density gradients in getting
information about the quantum critical behavior of the PP
transitions. In addition, the effects of contact misalignments are
presented. Finally, in Section 4, we present a summary of the
results and the main conclusions of this paper.

\section{Density gradients}
\subsection{Introduction}

Density gradients in Hall bars, as well as other macroscopic
inhomogeneities such as {\em contact misalignments} etc., are long
standing experimental problems that are hard to deal with in
detail and difficult to control in general. The analysis that
follows is based on the assumption which says that the {\em
spatial extend} of macroscopic inhomogeneities is typically much,
much larger than any of the microscopic or mesoscopic length
scales in the problem of transport of two dimensional electrons.
Under these circumstances one can apply the results of ordinary
electrodynamics. This means that one can describe the transport
process by assuming phenomenological transport equations with
slowly varying, position dependent transport coefficients. The
general problem is actually a very old one\cite{inhomogeneity}
and, as is well known, the complications are very specific to the
magneto resistance measurements and usually do not arise in
ordinary problems of quantum transport, in zero magnetic field
$B$.

It is important to emphasize that our physical objectives have
nothing to do with, say, the mesoscopic type of ideas on sample
inhomogeneity\cite{percol} that are inspired by Landau level
systems with smoothly varying, long ranged potential fluctuations.
For the low mobility samples that are of interest to us,
semiclassical ideas based on the {\em percolation} picture only
enter the problem in a fictitious regime of extremely low $T$
where, as we mentioned before, the experiment on quantum
criticality has been completely destroyed, as a result of
inhomogeneities in the sample. Quite obviously, the physical
objectives of scaling, at finite $T$, simply demand that one
starts the analysis from the opposite limit where the effects of
macroscopic inhomogeneities are negligibly small. Indeed, one of
the most important and difficult tasks that one is faced with is
to make sure that the experiment is conducted in a regime in $T$
where the sample inhomogeneities hardly affect the basic phenomena
of interest such that the appropriate corrections can be made.

Imagine a {\em local} density $\bf j$ of DC electrical current
that in principle may flow anywhere in the bulk of the sample as
long as certain local and macroscopic constraints are satisfied.
The {\em conservation of charge} is expressed by the continuity
equation ${\bf \nabla} \cdot {\bf j} = 0$. This implies that the
{\em total} electrical current $\bf J$ through the Hall bar is
constant and independent of the coordinate $x$ along the Hall bar
(Fig. 1). Next, the condition for having a {\em stationary state}
can be expressed using Maxwell's equation $\partial {\bf B} /
\partial t = {\bf \nabla} \times {\bf E} = 0$. Alternatively,
we may apply Stoke's theorem, $\oint {\bf E.dl}$ = 0, which says
that no source of voltage should exist inside the the Hall bar.

\begin{figure}
\begin{center}
\includegraphics[width=7cm]{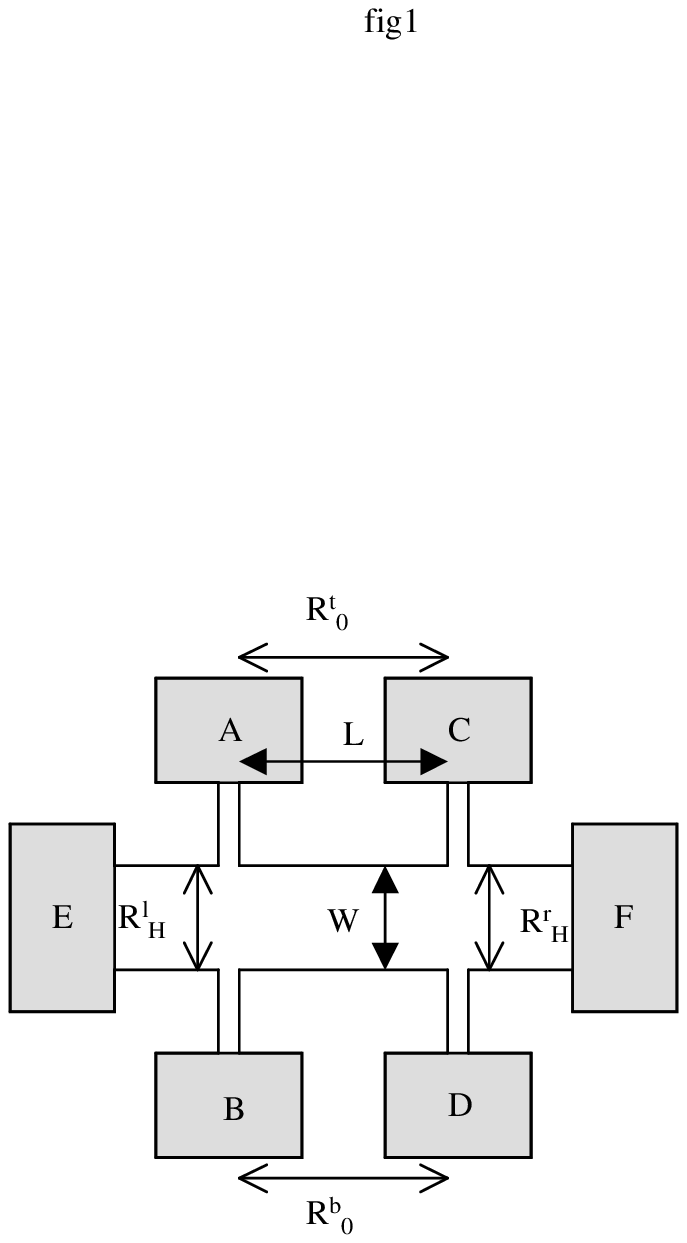}
\end{center}
\caption{Schematic diagram of Hall bar, L = 1mm, W = 0.5 mm }
\end{figure}

The phenomenological transport equations can be written in the
standard form

\begin{eqnarray}
E_{x} & = & ~~\rho_0 j_{x} - \rho_H j_y  \\
E_{y} & = & ~~ \rho_{H}j_{x} + \rho_0 j_y ,
\end{eqnarray}

Here $\rho_0$ and $\rho_H$ denote the {\em longitudinal} and {\em
transversal} or {\em Hall} resistivity respectively which may be
taken as phenomenological parameters that vary slowly in $x$ and
$y$. The aforementioned constraints imply that current density
${\bf j} =(j_x ,j_y )$ satisfies the following equations

\begin{eqnarray}
{\bf \nabla} \cdot {\bf j} & = & 0 \\
\rho_0 {\bf \nabla} \times {\bf j} & = & (\partial_y \rho_0 -
\partial_x \rho_H ) j_x - (\partial_x \rho_0 +
\partial_y \rho_H) j_y .
\end{eqnarray}
By writing ($j_{x}$ , $j_{y}$) = ($\partial_y \phi$ , $-\partial_x
\phi $) we can also express Eqs (3) and (4) in terms of a
differential equation for the scalar field $\phi$ alone,

\begin{equation}
{\rho_0 } {\Delta\phi} = - \left[ ( \partial_x \rho_0 +
\partial_y \rho_H ) \partial_x + (\partial_y \rho_0  - \partial_x \rho_H )
\partial_y  \right] \phi.
\end{equation}

\subsection{Linear approximation}
\subsubsection{Formalism}
Except for the very special cases to be discussed below it is not
only difficult to solve Eq. (5) in general but also very hard to
extract relevant and physical information by pursuing a purely
analytic approach to the problem. It is important to keep in mind,
however, that we are primarily interested in the effects that are
induced by {\em weak} macroscopic inhomogeneities. An adequate
description of the problem is obtained by employing what we call
the {\em linear approximation}. This amounts to having transport
parameters $\rho_0$ and $\rho_H$ that vary linearly in the
coordinates $x$ and $y$ according to

\begin{eqnarray}
\rho_0 & = & \rho_0^0 (1 + \alpha_x x + \alpha_y y )  \\
\rho_H & = & \rho_H^0 (1 + \beta_x x + \beta_y y ) .
\end{eqnarray}
Here, the $\alpha's$ and $\beta's$ are given as phenomenological
parameters that may vary with $T$ and the external field $B$ etc.
We assume for simplicity a rectangular geometry of length $L$
(x-direction) and width $W$ (y-direction, see Fig. 1). The
coordinates $x$ and $y$ are defined relative to the center of the
Hall bar. It is easy to see that a spatially varying current
density of the form

\begin{eqnarray}
j_x & = & j_x^0 (1+ a_y y) \nonumber \\
j_y & = & 0
\end{eqnarray}
satisfies all the aforementioned constraints and boundary
conditions. Working to linear order in the coordinates $x$ and $y$
one can write the transport equations as follows

\begin{eqnarray}
E_x & = & \rho_0^0 \left[ 1 + \alpha_x x + (\alpha_y + a_y ) y \right] j_x^0  \\
E_y & = & \rho_H^0 \left[ 1 + \beta_x x + (\beta_y + a_y ) y
\right] j_x^0 .
\end{eqnarray}
Notice that the constraint ${\bf \nabla} \times {\bf E} = 0$ fixes
the parameter of the current density according to $a_y = -
\alpha_y + \frac{\rho_H^0}{\rho_0} \beta_x $ such that the
transport equations can be expressed in terms of the
phenomenological quantities $\alpha_{x,y}$ and $\beta_{x,y}$ alone

\begin{eqnarray}
E_x & = &  \rho_0^0 ~\left[ 1 + \alpha_x x  + (
\frac{\rho_H^0}{\rho_0} \beta_x ) y \right] j_x^0
\\
E_y & = &  \rho_H^0 \left[ 1 + \beta_x x + (
\frac{\rho_H^0}{\rho_0} \beta_x  + \beta_y  - \alpha_y ) y \right]
j_x^0 .
\end{eqnarray}

These expressions can be used to compute the voltage drop $V_x =
\int_{-{L}/{2}}^{{L}/{2}} dx E_x$ that one measures on the {\em
top} and the {\em bottom} contacts of the Hall bar respectively.
Similarly one computes the potential difference $V_y =
\int_{-{W}/{2}}^{{W}/{2}} dy E_y$ that is measured on the {\em
right} and the {\em left} contacts of the Hall bar respectively.
The total current $J=\int_{-{W}/{2}}^{{W}/{2}} dy j_x =
j_x^0$. The results can be expressed in terms of the {\em
longitudinal resistances} $R_{0}^{t,b}$ that one measures on the
{\em top} and {\em bottom} of the Hall bar (Fig. 1)

\begin{equation}
R_{0}^{t}=\frac{L}{W}(\rho_{0}^0 +\rho_H^0 \beta_x
\frac{W}{2}),~~~~~~ R_{0}^{b}=\frac{L}{W}(\rho_{0}^0 -\rho_H^0
\beta_x \frac{W}{2})
\end{equation}
Similarly, we obtain the {\em Hall resistances} $R_{H}^{l,r}$ that
are being measured on the {\em left} and the {\em right} contacts
of the Hall bar,

\begin{equation}
R_{H}^{l}= \rho_{H}^0 (1 - \beta_x \frac{L}{2}),~~~~~~ R_{H}^{r}=
\rho_{H}^0 (1 +\beta_x \frac{L}{2}) .
\end{equation}
We see that the measurements on the Hall bar are affected by the
quantity $\rho_H^0 \beta_x$ only, i.e. the gradient of the Hall
resistance along the direction of the current.

\subsubsection{Reflection symmetry}

One immediately recognizes that the measured quantities
$R_0^{t,b}$ and $R_H^{l,r}$ are related to one another by {\em
symmetries} that can easily be tested and observed in the
experiment. Notice that under a change in the {\em polarity} of
the magnetic field $B$ the quantity $\rho_H^0$ in Eqs (13) and
(14) changes sign. This is unlike the quantities $\rho_0^0$,
$\alpha_{x,y}$ and $\beta_{x,y}$ which are all even in $B$.
Therefore, gradients in the electron density manifest themselves
primarily through the following symmetry

\begin{eqnarray}
B & \rightarrow & -B \\
R_0^t , R_0^b & \rightarrow & R_0^b , R_0^t \\
R_H^l , R_H^r & \rightarrow & - R_H^l , - R_H^r .
\end{eqnarray}

\subsubsection{Conclusion}
In as far as one can trust the {\em linear approximation}, a
'best' estimate for the local resistivity components $\rho_0$ and
$\rho_H$ is obtained by taking the {\em sum} of the measured
quantities on the {\em top} and {\em bottom} of the Hall bar and
those on the {\em left} and {\em right} hand sides respectively.
\begin{eqnarray}
\frac{R_{0}^{t} + R_{0}^{b}}{2} & = & \frac{L}{W} \rho_0^0
\\ \frac{R_{H}^{r} + R_{H}^{l}}{2} & = & \rho_H^0  .
\end{eqnarray}
At the same time, the {\em difference} between the various
measured quantities provides a numerical estimate for the
macroscopic inhomogeneities in the sample. Specifically,

\begin{equation}
R_{0}^{t}-R_{0}^{b} = R_{H}^{r}-R_{H}^{l} = \rho_H^0 \beta_x L
=\frac{\partial\rho_H}{\partial n} \delta n.
\end{equation}
Here, $\delta n$ denotes the typical difference between the local
densities $n$ of the electron gas that exist on the left and right
hand sides of the Hall bar respectively.

Several comments are in order. First of all, for a more detailed
understanding of the problem, it is important to know how and to
what extend the results of Eqs (18) and (19) are modified by the
presence of {\em non-linear} effects in the density gradients,
represented by the higher order terms in the series of Eqs (6) and
(7). Progress along these lines will be reported elsewhere where
we embark on a detailed quantitative analysis of the experimental
data, in particular those on the PP transition. Here we just
mention the fact that the results of this Section can be
systematically extended to include the higher order expansion
schemes, denoted by {\em quadratic approximation}, {\em cubic
approximation} etc.

In this paper our objectives are slightly different. To understand
the different ways in which the inhomogeneities manifest
themselves we shall, in what follows, discuss the the various
different regimes in $B$ separately. In Section 2.3 below we address
the {\em semiclassical} theory of transport and introduce an
exactly solvable model for density gradients. This enables us to
discuss the general features of inhomogeneity that one expects to
be relevant for the experimental studies at weak values of $B$.

In Section 2.4 we address {\em quantum Hall regime}. It turns out
that the PI transition, unlike the PP transition, lends itself to
an analytic study on inhomogeneity effects. The results of this
Section are particularly illuminating for a general understanding
of the experimental differences that exist between the PP and
PI transitions.

\subsection{Semiclassical regime}
\subsubsection{Introduction}
The transport problem at weak values of $B$ is described, as is
well known, by the Drude-Zener-Boltzmann theory. In the absence of
any weak quantum interference effects we have

\begin{equation}
\rho_0 = \frac{1}{\sigma_{cl}} ;~~ \rho_H = \frac{\omega_c
\tau}{\sigma_{cl}} .
\end{equation}
Here $\omega_c$ is the cyclotron frequency, $\omega_c =
\frac{eB}{m^*}$, and $\sigma_{cl}$ is given by the well known
expression

\begin{equation}
\sigma_{cl} = \frac{n e^2 \tau}{m^*} .
\end{equation}
To be able to study the effects of macroscopic density gradients
beyond the limitations of the {\em linear approximation} we
consider the special case where the electron density varies
exponentially according to

\begin{equation}
n (x,y) = n_0 e^{- a_x x - a_y y} .
\end{equation}
Here $x, y$ denote, as before, the coordinates relative to the
center of the Hall bar. Under these circumstances one can express
Eq. (5) in terms of the phenomenological parameters $a_x$ and $a_y$
alone,

\begin{equation}
\Delta \phi = -[(a_x + \omega_c \tau a_y ) \partial_x + (a_y -
\omega_c \tau a_x ) \partial_y ] \phi .
\end{equation}
The most important feature of this theory is that it can be solved
exactly.

\subsubsection{Exact solution}

The solution for $\phi$ which has the appropriate boundary
conditions can be written as follows

\begin{equation}
\phi = \phi(y) = - \frac{j_x^0}{a_y - \omega_c \tau a_x} e^{-( a_y
- \omega_c \tau a_x )y } .
\end{equation}
This result implies the following simple expression for the
current density
\begin{equation}
j_x = j_x (y) = {j_x^0} e^{-( a_y - \omega_c \tau a_x )y } , ~~j_y
= 0 .
\end{equation}
One can next proceed along the same lines as outlined before and
compute the the macroscopic magneto resistances $R_0$ and $R_H$ as
measured on a rectangular sample of size $L \times W$. The result
can be written as follows

\begin{eqnarray}
R_0^{t,b} & = & \frac{L}{W} \rho_0^0 \times e^{\pm \omega_c \tau
a_x \frac{W}{2}} \times
\nonumber \\
& \times &  \frac{\sinh (a_x \frac{L}{2})}{a_x \frac{L}{2} }
\times \frac{(a_y -\omega_c \tau a_x ) \frac{W}{2} }{ \sinh ((a_y
- \omega_c \tau a_x ) \frac{W}{2})}  \\
R_h^{r,l} & = & \rho_H^0 \times e^{\pm a_x \frac{L}{2}} \times
\nonumber \\
& \times & \frac{ \sinh (\omega_c \tau a_x \frac{W}{2})}{\omega_c
\tau a_x \frac{W}{2} } \times \frac{(a_y -\omega_c \tau a_x )
\frac{W}{2}}{ \sinh ((a_y - \omega_c \tau a_x ) \frac{W}{2})} .
\end{eqnarray}
Here, the quantities $\rho_0^0$ and $\rho_H^0$ are formally
defined at the center of the Hall bar, $x=y=0$, where the density
is given by $n(x,y) = n_0$. Specifically we have

\begin{equation}
\rho_0^0 = \frac{1}{\sigma_{cl}^0} ;~~ \rho_H^0 = \frac{\omega_c
\tau}{\sigma_{cl}^0} ,~~ \sigma_{cl}^0 = \frac{n_0 e^2 \tau}{m^*}.
\end{equation}
One readily verifies that Eqs (27) and (28), if expanded to lowest
order in a series in powers of $a_x$ and $a_y$, produces the same
results as those obtained before, in the so-called {\em linear
approximation}. However, the present results also provide a more
general insight into the structure of the theory that cannot be
obtained by focussing on the {\em linear approximation} alone. For
example, Eqs (27) and (28) indicate that {\em reflection
symmetry}, Eqs (15-17), is in fact a specific feature of the
{\em linear approximation} that in general is violated by the
higher order terms in the series.

\subsubsection{Beyond the linear approximation,~$\omega_c \tau \lesssim 1$}
Important for our present purposes are the following {\em general}
conclusions that are very useful in experimental studies of
macroscopic sample inhomogeneities. First, on the basis of Eqs
(27) and (28) one can write down the magneto resistance parameters
in terms of a series in powers of the dimensionless quantity
$\omega_c \tau$. The result takes the following form
\begin{eqnarray}
R_0^{t,b} & = & \frac{L}{W} \rho_0^0 \times \left\{\alpha_0 +
\omega_c
\tau \alpha_1^{t,b} + (\omega_c \tau)^2 \alpha_2 +\cdot\cdot\right\}\\
R_h^{r,l} & = & ~~\rho_H^0 \times \left\{ \beta_0^{r,l} +
\omega_c \tau \beta_1 ~+ (\omega_c \tau)^2 \beta_2 +\cdot\cdot\right\}
\end{eqnarray}
Generally speaking, the coefficients $\alpha_i$ and $\beta_i$ in
these series appear as complicated expressions, describing in
detail how the theory depends on the effects of macroscopic
density variations. For the specific example at hand one can
express the parameters $\alpha_i$ and $\beta_i$ in terms of a
series in powers of the phenomenological parameters $a_{x,y}$, Eqs
(27) and (28). The results, up to quadratic order in $a_{x,y}$,
are as follows

\begin{eqnarray}
\alpha_0 ~& = & 1+~~~~~~~~~~\frac{1}{6}  a_x^2 ( \frac{L}{2})^2
-\frac{1}{6} a_y^2 ( \frac{W}{2})^2
\\
\alpha_1^{t,b} & = & ~\pm a_x \frac{W}{2} + \frac{1}{3} a_x a_y
(\frac{W}{2})^2 \\
\alpha_2 ~& = & ~~~~~~~~~~~~~~~\frac{1}{3} a_x^2 ( \frac{W}{2})^2 \\
\nonumber \\
\beta_0^{r,l} & = & 1 \pm a_x \frac{L}{2} -\frac{1}{6} a_y^2
(\frac{W}{2})^2
\\
\beta_1 ~& = &  ~~~~~~~~~~~~~~~\frac{1}{3} a_x a_y
(\frac{W}{2})^2 \\
\beta_2 ~& = & ~~~~~~~~~~~~~~~~O(a^3 ).
\end{eqnarray}
Secondly, it should be mentioned that Eqs (30) and (31) still do
not capture the most general features of density gradients in the
semiclassical regime. The most general expression for the Hall
resistance also contains contributions that do not vanish in the
limit where $\omega_c \tau \rightarrow 0$\cite{unpub}.

We summarize the results of this Section as follows. Assuming that
the experiment at weak $B$ is not dominated by such phenomena like
{\em weak quantum interference}, {\em contact misalignments} etc.,
then the effects of sample inhomogeneity can generally be
classified into two distinctly different sectors, the {\em Linear}
effects and the {\em Nonlinear} effects. \vskip 2mm

\noindent{$\bullet$} {\em Linear effects.} These refer to the {\em
differences} in the measured resistances

\begin{equation}
R_0^t - R_0^b = R_H^r - R_H^l \approx \rho_H^0 \frac{\delta n}{n}.
\end{equation}
Here, $\frac{\delta n}{n}$ denotes the relative difference in the
electron density that generally exists near the opposite contact
pairs $AB$ and $CD$ of the Hall bar (see Fig. 1). \vskip 2mm

\noindent{$\bullet$} {\em Nonlinear effects.} These are observed
in the {\em sum} of the measured resistances

\begin{eqnarray}
\frac{R_0^t +R_0^b}{2} & = & \frac{L}{W} ~\rho_0^0 \times F_0
(\omega_c \tau )
\\
\frac{R_H^r +R_H^l}{2} & = & \gamma_0 + \rho_H^0 \times F_H
(\omega_c \tau ) .
\end{eqnarray}
These quantities with $F_0 \approx 1$, $F_H \approx 1$ and
$\gamma_0 \approx 0$ generally provide the best estimate for
resistivity components $\rho_0^0$ and $\rho_H^0$. The phrase {\em
nonlinear} indicates, however, that the functions $F_0 (\omega_c
\tau )$ and $F_H (\omega_c \tau )$ are generally different from
unity whereas the quantity $\gamma_0$ is generally different from
zero. From Eqs (30) and (31) we obtain explicitly

\begin{eqnarray}
F_0 (\omega_c \tau )= \alpha_0 +\omega_c \tau (\alpha_1^t
+\alpha_1^b )/2 + (\omega_c \tau )^2 \alpha_2 +\cdot\cdot
\\
F_H (\omega_c \tau )= (\beta_0^r + \beta_0^l )/2 + \omega_c
\tau \beta_1 + ( \omega_c \tau )^2 \beta_2 +\cdot\cdot
\end{eqnarray}
For completeness we list the general expression for
$\gamma_0$\cite{unpub}

\begin{equation}
\gamma_0 =  \rho_0^0 \left[\alpha_x \alpha_y - \alpha_{xy} \right]
\frac{LW}{12} .
\end{equation}
Here, the coefficient $\alpha_{xy}$ is phenomenological parameter
that - just like the $\alpha_{x}$ and $\alpha_{y}$ - appears in
the series of Eq. (6). It is readily verified that $\gamma_0$
vanishes for the specific problem at hand, as it should be.

As a final remark we can say that the {\em nonlinear} effects are
weaker than the {\em linear} effects by one order of magnitude
in $\frac{\delta n}{n}$. Nevertheless, the results give rise to a
$B$ dependence in the experimental data that is distinctly
different from what one expects on the basis of the semiclassical
theory of $\rho_0^0$ and $\rho_H^0$ alone.

\subsubsection{Beyond the linear approximation,~$\omega_c \tau \ggg 1$}
Next, we elaborate on the semiclassical results, Eqs (27) and
(28), for large values of $B$. Notice that in this limit the
Drude-Zener Boltzmann theory (Eq. 21) indicates that the electron
gas is in an {\em insulating} state. In terms of the conductivity
parameters $\sigma_{xx} = \frac{\rho_0}{\rho_0^2 + \rho_H^2}$ and
$\sigma_{xy} = \frac{\rho_H}{\rho_0^2 + \rho_H^2}$ one has
$\sigma_{xx} = \sigma_{xy} =0$ which is also known as the regime
of the {\em Hall insulator}. In this interesting regime one
observes, in the experiment at low $T$, some of the strongest
quantum features of the electron gas and, hence, the most dramatic
departure from semiclassical transport theory. The results of Eqs
(27) and (28) nevertheless provide an interesting and instructive
example of highly nonlinear effects due to density gradients.

The first thing to notice is that the field $B$ in Eqs (27) and
(28) only enters through the combination $\omega_c \tau a_x$. This
indicates that {\em homogeneous} ($a_x =0$) and {\em
inhomogeneous} ($a_x \neq 0$) systems behave very differently in
the limit $B \rightarrow \infty$. As an example we consider the
expression for the current density, Eq. (26). If we take the limit
$\omega_c \tau \rightarrow \infty$ while keeping $a_x$ at a finite
value then, dependent on the sign of $\omega_c \tau a_x$, we can
represent Eq. (26) as follows
\begin{equation}
j_x = j_x (y) = J ~\delta(y \mp \frac{W}{2} ) ,~~ j_y =0 .
\end{equation}
Here, the $\mp$ sign indicates that all the current is now
accumulated on either the {\em top} edge of the Hall bar or the
{\em bottom} edge. We have adjusted the amplitude $j_x^0$ in Eq.
(26) in such a way that $J = \int dy j_x$, the total current
through the system, remains finite.

Next, depending on whether the current flows along the {\em top}
edge or {\em bottom} edge, the expression for longitudinal
resistances becomes either
\begin{equation}
R_0^t = |\delta \rho_H^0 | ,~~ R_0^b =0 ,
\end{equation}
or

\begin{equation}
~~~~~~ R_0^t =0 , ~~~~~~R_0^b = |\delta \rho_H^0 | .
\end{equation}
Here, $\delta \rho_H^0 = -\rho_H^0 \frac{\delta n}{n}$ where
$\frac{\delta n}{n} = -a_x L$ denotes the relative
uncertainty in the density along the x-direction. Notice that
$\delta \rho_H^0$ is the same quantity that enters into the
expressions of {\em linear approximation}. Finally, the result for
the Hall resistances is the same in both cases and given by

\begin{equation}
R_H^r = R_H^l = \rho_H^0 .
\end{equation}
In deriving these expressions we have retained the terms to lowest
order in $a_x$ and $a_y$ only.

The results of this Section indicate that the strong $B$
predictions of the Drude-Zener-Boltzmann theory are strongly
affected, in a highly non-perturbative fashion, by the presence of
even weak inhomogeneities in electron density. Notice, however,
that under these extreme circumstances the theory nevertheless
retains {\em reflection symmetry}, Eqs (15-17). The results of
this Section actually provide a lucid example of a more general
statement which says that {\em reflection symmetry} is in fact a
quite commonly observed (albeit approximate) feature of the
magneto resistance data taken from the Hall bar, in the entire
range of experimental $B$. We shall return to this issue at a
later stage, in the experimental Section of this paper.

\subsection{Quantum criticality}
\subsubsection{PP transition}
It is easy to see that the macroscopic inhomogeneities are likely
to complicate the experiments on the PP transitions in the quantum
Hall regime. Notice that a spatially varying {\em electron
density} $n$ has roughly the same meaning as a spatially varying
{\em filling fraction} $\nu$ of the Landau level system, $\nu =
{n}/{n_B}$ where $n_B = eB/h$ denotes the degeneracy of the
Landau level. The typical corrections described in the previous
Section can be written as

\begin{equation}
\rho_H^0 \beta_x = \frac{\partial \rho_H}{\partial \nu}
\frac{\partial \nu}{\partial x} \propto T^{-\kappa} .
\end{equation}
Here we have used the fact that ${\partial \rho_H}/{\partial \nu}$
or ${\partial \rho_H}/{\partial B}$ is proportional to
$T^{-\kappa}$ which tends to diverge at the quantum critical point
(center of the Landau band) as $T$ approaches zero. This
complication of the PP transition has not been recognized
previously but it obviously throws fundamental doubts on the
accuracy of the previously accepted experimental value for
critical index $\kappa = 0.42$.

We next discuss the consequences for the PP transitions in some
more detail. To start, we recall that the transport parameters
with varying $\nu$ and $T$ depend on a single scaling variable $X$
only
\begin{eqnarray}
\rho_0 (\nu, T) & = & \rho_0 (X) ,\\
\rho_H (\nu, T) & = & \rho_H (X) ,
\end{eqnarray}
where

\begin{equation}\label{X}
X = \frac{\nu-\nu^*}{\Delta \nu (T)} = \frac{\nu -\nu^* } {\left(
\frac{T}{T_0} \right)^{\kappa}} .
\end{equation}
Here, $\nu^*$ denotes the {\em critical filling fraction} of
Landau level which is close to a half-integer, ${\Delta \nu (T)} =
{\left( \frac{T}{T_0} \right)^{\kappa}}$ describes the {\em width}
of the quantum critical phase transition with varying $T$, and $T_{0}$ is a
phenomenological parameter. Assuming the filling faction $\nu$ of
the Landau level system to be spatially varying

\begin{equation}
\nu = \nu_0 +\nu_x x + \nu_y y ,
\end{equation}
then Eqs (13) and (14) can be written as
\begin{eqnarray}
R_{0}^{t,b} & = & \frac{L}{W} \left( \rho_{0}^0 (X) \pm
\frac{W}{L}
(\partial_X \rho_H^0 ) \frac{\delta \nu_x}{\Delta \nu(T)} \right), \\
R_{H}^{r,l} & = & ~~~~~~\rho_{H}^0 (X) \pm ~~~~(\partial_X
\rho_H^0 ) \frac{\delta \nu_x}{\Delta \nu(T)} .
\end{eqnarray}
Here,

\begin{eqnarray}
{\delta \nu_x} & = & \nu_x \frac{L}{2} ,
\nonumber \\
{\delta \nu_y} & = & \nu_y \frac{W}{2}
\end{eqnarray}
represent the uncertainties in the filling fraction due to the
density variations in the $x$ and $y$ directions respectively.

On the basis of Eqs (53) and (54) one readily derives the
following qualitative features of the magneto resistance data that
are generally observed in the experiment (Section 3.3).

\noindent{\em $~\bullet~$ Hall resistance~~} First, the result for
the Hall resistance Eq. (54) describes the lowest order terms of a
Taylor series expansion. This indicates that in practice the
inhomogeneities in the electron density manifest themselves
through the following behavior of the Hall resistances

\begin{equation}
R_{H}^{r,l}  =  \rho_{H} (X^{r,l} ) , ~X^{r,l} = \frac{\nu_0
-\nu^* \pm \delta \nu_x }{\Delta \nu (T)} .
\end{equation}
In words, the quantities $R_H^{r,l}$ that are taken from the {\em
right} and the {\em left} hand sides of the Hall bar probe, in
effect, different values of the critical filling fraction, $\nu^*
- \delta \nu_x$ and $\nu^* + \delta \nu_x$ respectively.

More precisely formulated, one still has to express a spatially
varying filling fraction $\nu$, Eq. (52), in terms of a fixed,
spatially varying electron density $n$. Write

\begin{equation}
n = n_0 ( 1 + \eta_x x + \eta_y y),
\end{equation}
then the quantities $\nu_0$, $\nu_x$ and $\nu_y$ are expressed in
terms of the fixed quantities $n_0$, $\eta_x$ and $\eta_y$ as
follows

\begin{equation}
\nu_0 = \frac{n_0}{n_B} , ~~\nu_x = \nu_0 \eta_x , ~~ \nu_y =
\nu_0 \eta_y .
\end{equation}
This suggests a slightly different expression for $X^{r,l}$,
\begin{equation}
X^{r,l} = \frac{\nu_0 -\nu_{r,l}^* }{\left( \frac{T}{T_0^{r,l}}
\right)^\kappa } ,
\end{equation}
where besides a difference in $\nu^*$ there is also a difference
in temperature scale $T_0$

\begin{equation}
\nu_{r,l}^* = \nu^* (1 \mp \eta_x \frac{L}{2} ) , ~~T_0^{r,l} =
T_0 (1 \pm \kappa^{-1} \eta_x \frac{L}{2} ) .
\end{equation}
\noindent{\em $~\bullet~$ Longitudinal resistance~~} The situation
for the longitudinal resistances $R_0^{t,b}$ is very different.
Notice that the derivative $\partial \rho_H /
\partial X$ with varying $X$ looks very much the same as the
function $\rho_0 (X)$. Therefore one expects the following
qualitative behavior of the longitudinal resistances

\begin{equation}
R_{0}^{t,b}  \approx  \frac{L}{W} A^{t,b} \rho_{0} (X) ,
\end{equation}
i.e. the $R_0^{t,b}$ data, taken from the {\em top} and the {\em
bottom} contacts of the Hall bar, mainly display difference in
amplitude, $A^{t} \neq A^{b}$.

\noindent{\em $\bullet$~Experimental criterion~} Finally, Eqs (53)
and (54) define the following experimental condition

\begin{equation}
\frac{\delta\nu_{x,y}}{\Delta\nu (T)} \ll 1 ~~~~~~~(PP)
\end{equation}
This indicates that in order to be able to extract reliable
information from the experimental data on the PP transition, the
uncertainties $\delta\nu_{x,y}$ in the filling fraction due to
sample inhomogeneities should be small as compared to the actual
width $\Delta\nu (T)$ of the quantum critical phase. As we already
mentioned earlier, this condition is not trivially satisfied in
general. This means that Eqs (18) and (19) simply do not provide a
good estimate for the transport parameters of the PP transition
and nonlinear effects generally become important.

\subsubsection{PI transition}

The experimental situation of the PI transition is very different.
We proceed by quoting some of the principal new results as
reported in Ref.\onlinecite{pruisken:stretch}. The explicit form of the
scaling functions in this case is given by

\begin{eqnarray}
\rho_0 (\nu, T) & = & \rho_0 (X) = e^{ -X} ,\\
\rho_H (\nu, T) & = & \rho_H (X) = 1 .
\end{eqnarray}
The most important feature of these results is that the Hall
resistance at low $T$ remains quantized ($\rho_H (X) = 1$) inside
the quantum critical phase $|X| \lesssim 1$. This means that the
experimental transport data are {\em unaffected} by the presence
of macroscopic inhomogeneities in the sample, at least within the
{\em linear approximation}. This unexpected feature of the lowest
Landau level clearly suggests that, for a given sample, the
experiment conducted on the PI transition is generally more
reliable than those conducted on the PP transitions.

The PI transition is one of the very rare examples where the
inhomogeneity problem can be solved exactly. Using Eqs (63) and
(64) we obtain for Eq. (5)

\begin{equation}
\Delta \phi  =  \left[ \alpha_x \partial_x + \alpha_y
\partial_y \right] \phi ,
\end{equation}
where

\begin{eqnarray}
\alpha_x & = & \alpha_x (T) = \frac{\nu_x }{\Delta \nu(T)} ,
\nonumber \\
\alpha_y & = & \alpha_y (T) = \frac{\nu_y }{\Delta \nu(T)}
\end{eqnarray}
are spatially independent quantities. The general solution is
readily obtained by a separation of variables. A solution with the
appropriate boundary conditions can be written as follows
\begin{equation}
\phi (x,y)  = \frac{j_x^0}{\alpha_y} \exp \{ \alpha_y y \}
\end{equation}
such that the expression for the current density becomes

\begin{equation}
j_x = j_x (y) =j_x^0 \exp \{\alpha_y y \} , ~~~j_y =0 .
\end{equation}
Proceeding along the same line as in the previous Section we
obtain the macroscopic resistances as follows

\begin{eqnarray}
R_{0}^{t,b} & = & \frac{L}{W} \rho_0^0 \times ~\frac{{\delta
\nu_y}\sinh \left( \frac{\delta \nu_x}{\Delta\nu (T)}
\right)}{{\delta\nu_x} \sinh \left( \frac{\delta \nu_y }{\Delta\nu
(T)} \right)}
\nonumber \\
& \approx & \frac{L}{W} \rho_0^0 \times \left[1+ \frac{1}{6} (
\frac{\delta \nu_x}{\Delta\nu(T)})^2 -
\frac{1}{6} (\frac{\delta \nu_y}{\Delta\nu(T)})^2 \right]\\
R_{H}^{r,l} & = &  1
\end{eqnarray}
indicating that the two measurements of $R_0$ and those of $R_H$
are now the same.

\noindent{\em $\bullet$~Experimental criterion~} Notice that the
condition

\begin{equation}
\frac{1}{6} \left( \frac{\delta \nu_{x,y}}{\Delta\nu (T)}
\right)^2 << 1 ~~~~~~~(PI)
\end{equation}
defines the regime in $T$ where the experiment on the PI
transition is unaffected by the macroscopic inhomogeneities. This
condition is generally much weaker than the one that is imposed on
the PP transition, Eq. (62).

These surprising aspects of the PI transition have all been
verified recently, by the experiment on a low mobility
$InP/InGaAS$ heterostructure at strong $B$\cite{pruisken:stretch}.
Quantum criticality was studied in the range $0.2 \lesssim T
\lesssim 4K$ with a quoted uncertainty in the filling fraction
$\frac{\delta \nu_{x,y}}{\Delta\nu(T)} \lesssim 0.2$ which is
consistent with Eq. (71).

\subsection{Universal scaling functions}
\subsubsection{Introduction}
As we mentioned earlier, the experiments on PI
transition\cite{pruisken:stretch} have led to a new estimate for
the critical exponent ($\kappa = 0.57$) which is slightly
different from the previously established one ($\kappa = 0.42$),
by H.P. Wei {\em et. al.}\cite{wei:prl88}. However, more than
providing a better estimate for critical exponents alone, the
results on the PI transition primarily indicate, for the first
time, that the {\em scaling functions} of the transport parameters
are {\em universal}. These new advances have fundamental
consequences, not only for the experiments on the PP transitions
but also for the quantum theory of conductances as a whole.

To understand the significance of Eqs (63) and (64) for the
experiments on the PP transitions we shall, in what follows,
briefly summarize some of the main predictions of the {\em
renormalization theory} of the quantum Hall
effect\cite{{pruisken:prl88},{paperIII},{EuroLett},{twoloop}}. In
the end this leads to a conceptual framework that will enable us
to address the problems associated with the PP transitions in much
greater detail.

\vskip 3mm

To start we convert, in a standard manner, the {\em resistivities}
of Eqs (63) and (64) into {\em conductivity} parameters
$\sigma_{xx}$ and $\sigma_{xy}$. However, for our present purposes
it is important to quote a slightly more general result for the PI
transition that involves two independent scaling variables $X$ and
$\eta$ \cite{pruisken:stretch}.

\begin{eqnarray}\label{sxx}
\sigma_{xx} (X,\eta) & = & \frac{e^{-X}}{1+ 2\eta e^{-X} + e^{-2X}} \\
\label{sxy}
\sigma_{xy} (X,\eta) & = & \frac{1+ \eta e^{-X}}{1+
2\eta e^{-X} + e^{-2X}} .
\end{eqnarray}
Here, $\eta = \eta(T)$ denotes the leading {\em irrelevant}
scaling variable

\begin{equation}\label{irrel}
\eta(T) = \pm \left( \frac{T}{T_1} \right)^{y_\sigma}
\end{equation}
with $T_1$ a phenomenological parameter and $y_\sigma =2.5$ the
leading {\em irrelevant} exponent. Notice that Eq. (\ref{sxy})
interpolates between the {\em quantum Hall plateau} $\sigma_{xy}
=1$ at large values of $X$ and the insulating phase with
$\sigma_{xy} =0$ that occurs when $X \rightarrow -\infty$. Eqs
(\ref{sxx}) and (\ref{sxy}), when plotted as $T$ driven flow lines
in the $\sigma_{xx}$, $\sigma_{xy}$ conductivity plane, provide a
lucid experimental demonstration of the scaling predictions made
by the {\em renormalization theory} of the quantum Hall effect
(see Appendix). Of general importance are two distinctly different
{\em symmetries} that control, to a large extend, scaling behavior
of the quantum Hall regime. These symmetries are

\vskip 3mm

\noindent{\em $\bullet$~Particle-hole symmetry.~~~} It is readily
verified that Eqs (\ref{sxx}) and (\ref{sxy}) satisfy the following
relations
\begin{eqnarray}\label{phxx}
\sigma_{xx} (X,\eta) & = &  \sigma_{xx} (-X,\eta)
\\ \label{phxy}
\sigma_{xy} (X,\eta) & = & 1 - \sigma_{xy} (-X,\eta) .
\\ \nonumber
\end{eqnarray}
This indicates that the line $\sigma_{xy} = \frac{1}{2}$ generally
appears as an axis of symmetry in the $\sigma_{xx}$, $\sigma_{xy}$
conductance plane.

\vskip 3mm

\noindent{\em $\bullet$~Periodicity in $\sigma_{xy}$.~~~}  Next,
there is the general statement which says that the scaling
behavior is periodic in $\sigma_{xy}$. This means that the scaling
results for adjacent quantum Hall plateaus $\sigma_{xy} =k$ and
$k+1$ can be obtained from Eqs (\ref{phxx}) and (\ref{phxy})
following the transformation (see Appendix)

\begin{eqnarray}\label{ll1}
\sigma_{xx} (X,\eta) & \rightarrow & \sigma_{xx}  (X^{(k)} , \eta^{(k)} ) \nonumber \\
\sigma_{xy} (X,\eta) & \rightarrow & \sigma_{xy} (X^{(k)} ,
\eta^{(k)} ) + k.
\end{eqnarray}
Here, $X^{(k)}$ and $\eta^{(k)}$ have the same meaning as the
original definitions except for a change in the critical filling
fraction $\nu^* \rightarrow \nu_k^*$ as well as the temperature
scales $T_0 \rightarrow T_0^{(k)}$ and $T_1 \rightarrow T_1^{(k)}$
which are determined by the detailed microscopic properties of the
electron gas. These quantities are in general different for
different Landau levels, and for different samples. More
specifically we have

\begin{eqnarray}\label{ll2a}
X_k & = & \frac{\nu - \nu_k^*}{\Delta^{(k)} \nu (T)} = \frac{\nu -
\nu_k^*}{\left( {T}/{T_0^{(k)}} \right)^\kappa } \\ \label{ll2b}
\eta_k & = & \pm \left( {T}/{T_1^{(k)}} \right)^{y_\sigma} .
\end{eqnarray}

\vskip 3mm
\subsubsection{Observability}
The quantities $T_0$ and $T_1$ define the range in $T$ where the
quantum critical phenomenon is observable. Since there are many
factors involved, in the actual value of these parameters, one
obviously needs the help of the renormalization theory to make
sure that the experiment is interpreted in the appropriate manner.
Generally speaking, one is faced with two major complications that
clearly explain why it is so difficult to fine tune the
experimental design such that the asymptotic $T \rightarrow 0$
limit of scaling can be observed in the laboratory.

\noindent{\em $\bullet$ ~~Weak localization~~~} First of all there
are the crossover effects due to {\em weak localization}. This
means that the actual value of $T_1$ may render arbitrary small
and very different phenomena are observed at finite $T$. For spin
polarized electrons there is the following generic expression for
$\sigma_{xx}$ in the regime of weak quantum
interference\cite{EuroLett,Finkel,twoloop}

\begin{equation}\label{weakloc}
\sigma_{xx} = \sigma_{xx}^0 + \beta \ln \frac{T}{T_f} .
\end{equation}
Here, $\sigma_{xx}^0$ denotes the semiclassical result, $\beta =
1/\pi$ a universal constant and $T_f= \frac{\sigma_{xx}^0}{16\pi
z_0 k_B}$ with $z_0$ the {\em singlet interaction amplitude} is
the typical $T$ scale for weak quantum interference
processes\cite{paperIII}. Eq. (\ref{weakloc}) is identified as the
asymptotic $\eta \rightarrow -\infty$ (or $\sigma_{xx}^0
\rightarrow \infty$) limit of Eq. (\ref{ll1}), i.e.

\begin{equation}\label{weakloc1}
\sigma_{xx} (X,\eta \rightarrow -\infty)  \approx  {\beta} \ln
\left( |\eta|^{1/y_\sigma} \right) ,
\end{equation}
where $\eta$ and $y_\sigma$ are defined as before, Eq.
(\ref{irrel}). Eqs (\ref{weakloc}) and (\ref{weakloc1}) imply that
the quantities $T_f$ and $T_1$ are related according to

\begin{equation}\label{t1}
T_1 = T_f ~e^{-\frac{\sigma_{xx}^0}{\beta}} ,
\end{equation}
indicating that the characteristic scale for observing {\em
quantum criticality}, $T_1$, vanishes {\em exponentially} as
$\sigma_{xx}^0$, i.e. the starting point for scaling, increases.
This typically happens in the semiclassical regime at weak $B$,
the center of the higher Landau bands of systems with
predominantly short ranged potential fluctuations but also in the
fractional quantum Hall regime, notably the half-integer effect.

\noindent{\em $\bullet$ ~~Long range impurity potentials~~~} Next,
there are the crossover effects due to {\em smooth} potential
fluctuations. These lead to arbitrary small values of the
parameter $T_0$. This is the main reason why quantum criticality
is in general not observed on samples that are used for, say,
fractional quantum Hall purposes. At finite $T$ one typically
observes a {\em linear} dependence on $T$\cite{shahar:ssc97}

\begin{equation}\label{semiclass}
\Delta\nu (T) = a+bT .
\end{equation}
This empirical result naively indicates that the {\em width}
$\Delta\nu (T)$ of the PP and PI transitions remains finite as $T$
approaches absolute zero. However, in Ref.\onlinecite{paperIII} it was shown
explicitly that Eq. (\ref{semiclass}) is a typical result of {\em
semiclassical transport} theory and much lower $T$ is generally
necessary before the algebraic behavior $\left( \frac{T}{T_0}
\right)^\kappa$ is observed. For this purpose a network with
percolating {\em edge states} has been considered where the length
scales (mean free path) associated with {\em inelastic} processes
are usually much shorter than those determined by the {\em
elastic} scattering events. The numerical value of the width
$\Delta\nu (T=0) =a$ was found to be strongly dependent on the
{\em range} $\lambda$ of the impurity potential and an elementary
analysis leads to $a = \frac{r_c^2}{\lambda^2}$ with $r_c$
denoting the magnetic length of the electron gas.

It is important to emphasize that the experimental complications
in observing quantum criticality have been addressed many times
before and at many places elsewhere. The findings of this paper
add as yet another major complication to the list. To summarize
the results we proceed by converting Eq. (77) back into
resistivities. We obtain the following general expressions for the
PP transitions ( dropping, for simplicity, the subscript $k$ on
the $X$ and $\eta$)

\begin{eqnarray}
\rho_0 (X,\eta)& = & \frac{e^{-X}}{(k+1)^2 + 2\eta k(k+1) e^{-X} +
k^2 e^{-2X}} ~~~~\\ \nonumber \\
\rho_H (X,\eta)& = & \frac{k+1+\eta (1+2k)
e^{-X} + k
e^{-2X}}{(k+1)^2 +2\eta k(k+1) e^{-X} + k^2 e^{-2X}}.~~~~\\
\nonumber
\end{eqnarray}

\noindent{\em $\bullet$~PI transition~~} We see that this general
result is rather complex in spite of the fact that the original
expressions for the PI transition ($k=0$) are quite simple.
Explicitly we have (compare Eqs (72) and (73))

\begin{equation}\label{rhoPP}
\rho_0 (X) = e^{-X} , ~~\rho_H (X,\eta) =1+\eta e^{-X}~~~~~~(k=0).
\end{equation}
Notice that this (more general) result clearly demonstrates the
fundamentally different physics that one can associate with the
{\em quantum critical phenomenon} on the one hand, and the {\em
ordinary quantum Hall effect} on the other. Whereas at the PI
transition ($X \approx 0$) the Hall resistance is exactly
quantized except for corrections that vanish {\em algebraically}
as $T$ goes to zero, in the ordinary plateau regime ($X
\rightarrow \infty$), however, the corrections are {\em
exponential} in $T$ (more precisely, one expects that the
exponential $e^{-X}$ in Eq. (\ref{rhoPP}), in the limit $X
\rightarrow \infty$, is replaced by an appropriate expression for
variable range hopping such as $e^{-\frac{2}{\sqrt{T\Gamma\xi}}}$
~\cite{hopping}). Eq. (\ref{rhoPP}) can therefore be considered as
one of the prettiest and most significant experimental tests of
scaling.

Finally, we specialize to the experimental differences between the
PI and PP transitions. From Ref.\cite{pruisken:stretch}
 we quote the following relation
between the experimentally observed {\em resistivity tensor}
$\rho_{i,j}$ (with $i$ and $j$ denoting the coordinates $x,y$) of
the PI transition, and the quantities of actual interest, $\rho_0$
and $\rho_H$
\begin{equation}
\rho_{ij} = S_{ij} \rho_0 (X) + \epsilon_{ij} \rho_H (X, \eta) .
\end{equation}
Here, the so-called {\em stretch tensor} $S_{ij}$ contains all the
imperfections of the Hall bar measurement due to density
gradients, contact misalignments etc. Assuming the geometry of a
parallelogram obtained by rotating a rectangle over a small angle
$\theta$ the result is

\begin{eqnarray}
S_{ij} (B,T) = \left[\begin{matrix}\frac{1}{cos\theta} &
-tg\theta~e^{\frac{\delta\nu_y}{\nu_0(T)}} \cr
tg\theta~e^{\frac{\delta\nu_x}{ \nu_0(T)}} & \frac{1}{cos\theta}
\end{matrix}\right]  . \label{eq:stretchtensor}
\end{eqnarray}
These results are valid as long as Eq. (71) is satisfied.

\noindent{\em $\bullet$~PP transition~~} As we have seen, the
results are very different in this case. Assuming that the main
effect comes from density gradients, then the quantity $\partial_X
\rho_H$ entering the linear approximation, Eqs (53) and (54), can
be obtained explicitly and the result is

\begin{equation}
\frac{\partial \rho_H}{\partial X} = - \rho_0 \left[
\frac{\rho_0}{\rho_0^{max}} +\eta (1- \frac{\rho_0}{\rho_0^{max}})
\right] .
\end{equation}
Here, $\rho_0^{max} = \rho_0^{max} (\eta)$ denotes the maximum
value of $\rho_0 (X,\eta)$ which is slightly different from the
critical resistivity $\rho_0^* (\eta)$ according to

\begin{eqnarray}
\rho_0^{max} (\eta) &=& \rho_0 (X= \ln \frac{k}{k+1} ,\eta )
\nonumber \\
&=& \frac{1}{2k(k+1)(1+\eta)}
\\ \nonumber \\
\rho_0^{*} (\eta) &=& \rho_0 (X= 0, \eta )
\nonumber \\
&=& \frac{1}{1 + 2 k(k+1)(1+\eta) } .
\end{eqnarray}

\section{Experimental design and results}
\subsection{Specifics of the sample}
The sample used in this work is a modulation doped $In_{0.7} Ga_{0.3} As$ 
quantum well of width $9 nm$ embedded in a $150 nm$ 
$InP$ buffer and $10 nm $ spacer grown by metalorganic
vapor phase epitaxy (MOVPE). On top of the spacer a $35 nm$ 
layer of $InP$ is grown with a $Si$ doping of $6 \times 10^{17}/cm^{3}$.

The numerical values for the electron density and mobility are
obtained from a measurement of the Hall resistance and
longitudinal resistance at low $B$. However, as an integral part
of our investigations we shall deal with the complications due to
macroscopic sample inhomogeneities which, according to the
previous Sections of this paper, is a highly non-trivial exercise
by itself. Keeping our principle objectives in mind we may report
the experimental and sample specific features as follows. The dark
value of mobility is $20,000 cm^{2}$ volt$^{-1}$ sec$^{-1}$ for an
electron density of $1.40 \times 10^{11}/cm^{2}$ at $4.2 K$.
Different values of the electron density, varying between $1.40
\times 10^{11}/cm^{2}$ and $4.66 \times 10^{11}/cm^{2}$, are
obtained by shining the light from a red $LED$ source, excited
with a $1 \mu A$ current, which is kept on for a controlled
duration. The enhanced electron density values are retained for
long times at low $T$ by persistent photoconductivity (PPC) which
confirms the good quality of the interface of the quantum well.
The electrical measurements are conducted at a frequency of $10
Hz$ using lock-in techniques. The electrical current ranges from
$10$ to $50 nA$, $T$ is varied between $2 K$ and $10 K$ and the
external field $B$ covers the range up to $8 T$.

\begin{figure}
\begin{center}
\includegraphics[clip, width=8.0 cm]{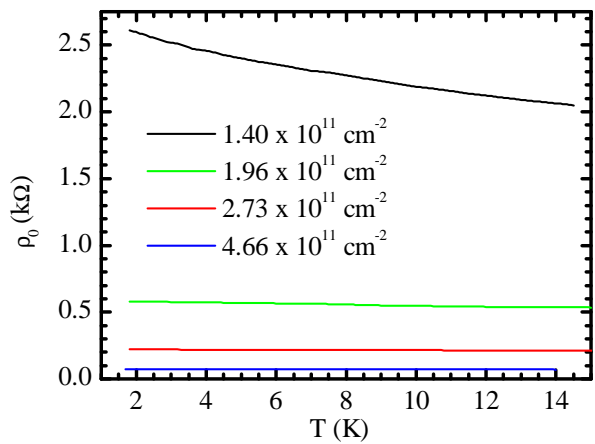}
\end{center}
\caption{\label{f2} Temperature dependance of resistivity at four
different densities }
\end{figure}

In Fig. 2 we plot the resistivity of the sample measured at $B=0$
and with varying values of $T$. Out of the four different values
of the electron densities that have been considered, only the data
taken at the lowest value of the density ($1.40 \times
10^{11}/cm^{2}$) display a significant enhancement of the
resistance as $T$ decreases. As the electron density is increased
up to the value $4.66 \times 10^{11}/cm^{2}$, the resistance
decreases by a factor of about 40 and remains almost constant over
the range of experimental $T$. As we shall discuss further below,
this dependence on $T$ indicates that {\em weak quantum
interference effects} are observable only in the experiment
conducted at the lowest value of the density. For the other
transport measurements at weak $B$, taken from the sample at the
three higher densities, these effects play hardly a role of
significance.

The Hall measurements at low $B$ show that the mobility of the
electron gas enhances with increasing values of the carrier
density in the range $(1.40 - 4.66) \times 10^{11}/cm^{2}$. This
enhancement, which is primarily the result of screening of the
potential fluctuations due to the static impurities and alloy
disorder \cite{screening}, indicates that for each value of the
electron density we are dealing with fundamentally different
impurity characteristics and, hence, with a truly different
sample.

\begin{figure}
\begin{center}
\includegraphics[clip, width=8.0 cm]{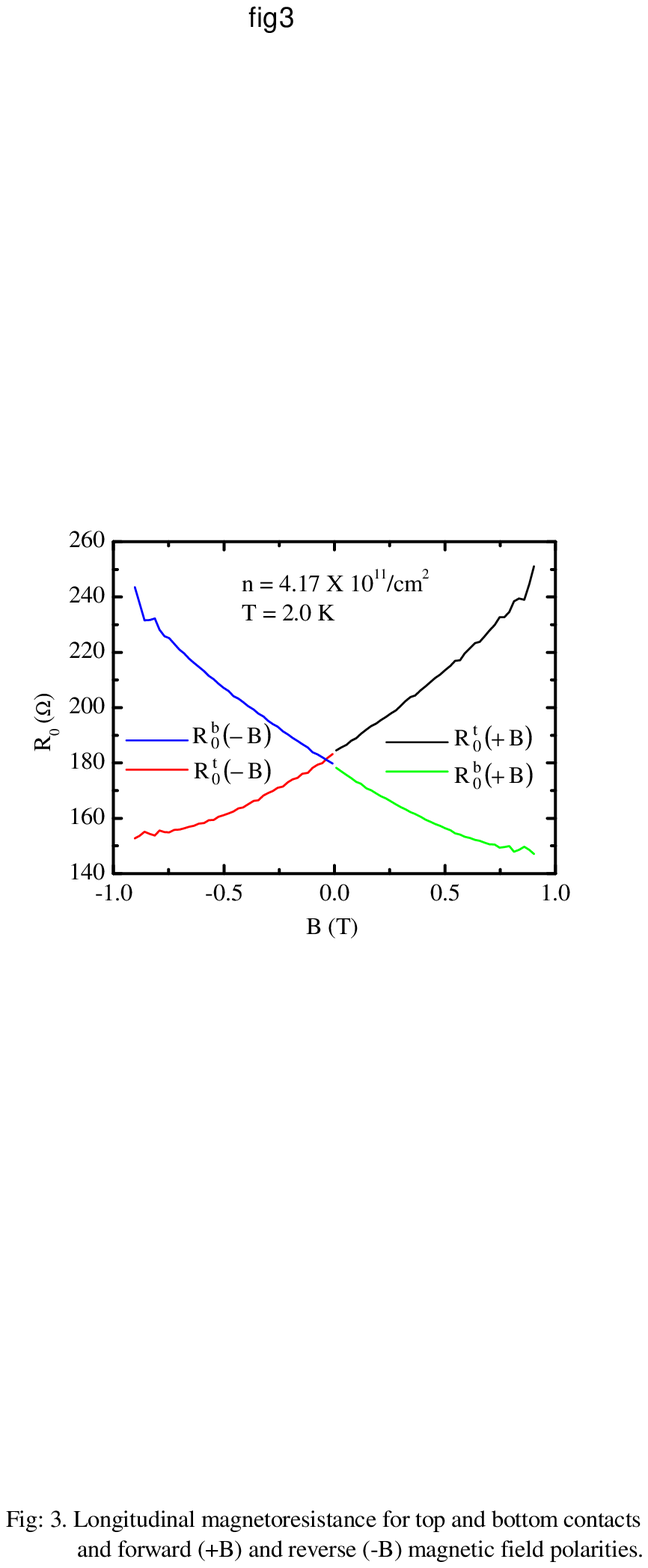}
\end{center}
\caption{\label{f3} Asymmetric longitudinal magnetoresistance }
\end{figure}

\subsection{Semiclassical regime}
\subsubsection{Aspects of symmetry}
\noindent{In} Fig. 3 we plot a typical example of the effect of
the polarity of the external field $B$ on the magneto resistance
data, $R_0^t$ and $R_0^b$, taken from the {\em top} and the {\em
bottom} pair of contacts of the Hall bar respectively. These
anomalous results for opposite polarities of the $B$ field clearly
display the symmetry

\begin{equation}
R_0^t (B) = R_0^b (-B) , R_0^b (B) = R_0^t (-B)
\end{equation}
that was predicted by the theory of {\em linear density
gradients}, Eqs. (15) and (16). Fig. 3 therefore permits us to apply the
theory of macroscopic inhomogeneities to the experimental data on
the semiclassical regime. In the end we shall compare the results
with those obtained from similar investigations of the quantum
Hall regime.

\subsubsection{Numerical estimate of density gradients}
Following Eq. (20) of Section 2 we obtain the {\em relative}
variation in the electron density of the sample according to

\begin{equation}
\frac{\delta n}{n} = 2\frac{R_H^r - R_H^l}{R_H^r + R_H^l} =6.2 \%
.
\end{equation}
It is important to emphasize that the measurements of $R_0^t -
R_0^b$ and $R_H^r - R_H^l$ are fundamentally the same and
therefore do not provide independent estimates for the quantity
$\frac{\delta n}{n}$.

\subsubsection{Nonlinear effects}
We next apply Eqs (39) and (40) in order to obtain the best
estimate for the local resistivity tensor. First we plot, in Fig.
4, the experimental data for $R_0^t + R_0^b$, normalized at $B=0$,
with varying values of $B$ and $T$ fixed at $2K$. The data for the
lowest value of the electron density, unlike those for the higher
densities, show clearly a negative magneto resistance. This result
is quite consistent with the statements made in the beginning
which say that {\em weak quantum interference} effects are mainly
displayed in the low density data and hardly significant at higher
values of the electron density. Notice, however, that
the data of Fig. 4, corresponding to the three higher densities,
nevertheless show a weak dependence on $B$ that cannot be
explained on the basis of the classical Drude-Zener-Boltzmann
result for $\rho_0$ alone. We attribute this weak $B$-dependence
in Fig. 4 to the {\em nonlinear} aspects of the density
inhomogeneities that have been introduced and discussed in Section
2, notably Eq. (39). This claim can be justified, at least in a
rough manner, if one compares the relative size of the {\em
linear} effects, like those shown in Fig. 3, with the relative
size of the {\em nonlinear} effects. On the basis of the discussion
given in Section 2.3 one expects the latter to behave like the
square of the former. This result qualitatively explains the
$B$-dependence as observed in the transport data at the three
higher densities, Fig. 4.

\begin{figure}[h]
\begin{center}
\includegraphics[clip, width=8.0 cm]{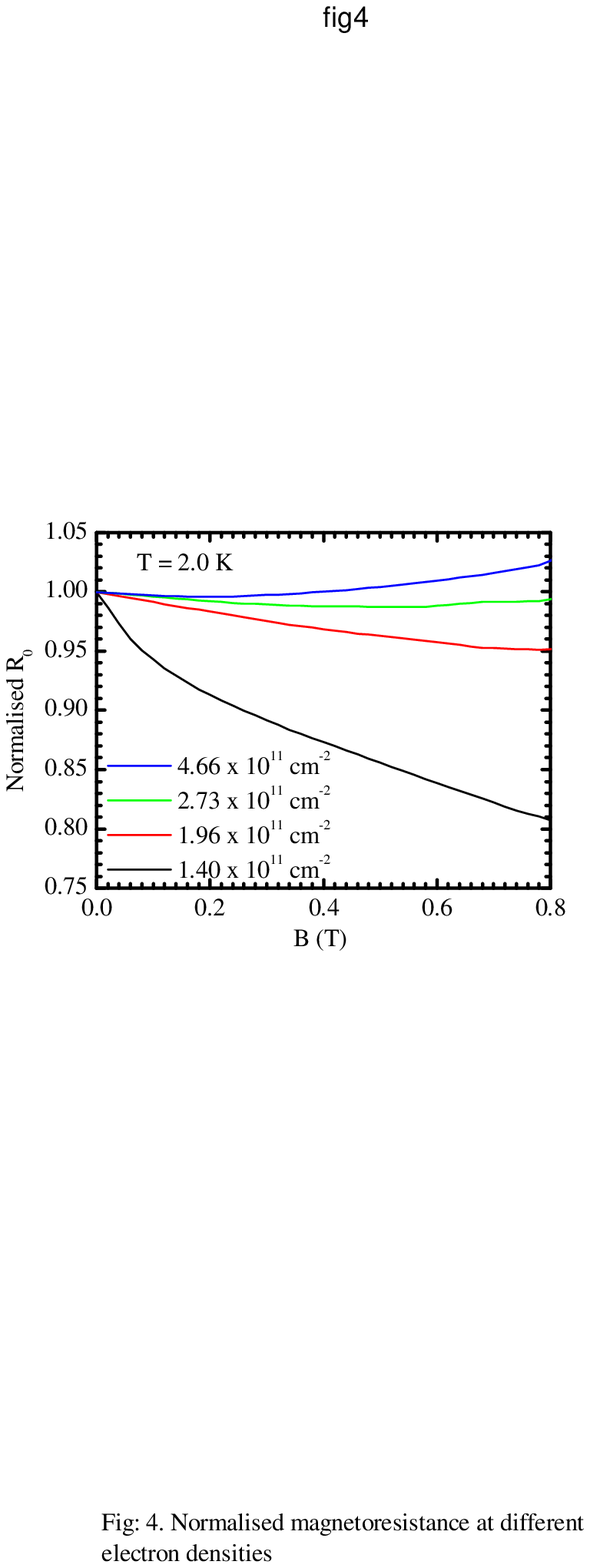}
\end{center}
\caption{\label{f4} Normalized magneto resistance versus $B$ at
$T=2K$. }
\end{figure}

In summary we can say that the effects of macroscopic
inhomogeneities, inherent to the experiment, are adequately
described and well accounted for by our theory of {\em linear} and
{\em nonlinear} density gradients, at least  as far as the
semiclassical regime of transport is concerned.

\subsubsection{The semicircle law}
As a next step toward a better understanding of the semiclassical
regime we next convert the experimental magneto resistance data
into a conductivity data using the standard formulae

\begin{equation}
\sigma_{xx} = \frac{\rho_0}{\rho_0^{2}+\rho_H^{2}},~~~~~ \sigma_{xy}
= \frac{\rho_H}{\rho_0^{2}+\rho_H^{2}} .
\end{equation}
where

\begin{equation}\label{r0h}
\rho_0 = \frac{W}{2L} (R_0^t + R_0^b ),~~~~~\rho_H =\frac{1}{2}
(R_H^r + R_H^l ).
\end{equation}

\begin{figure}[h]
\begin{center}
\includegraphics[clip, width=8.0 cm]{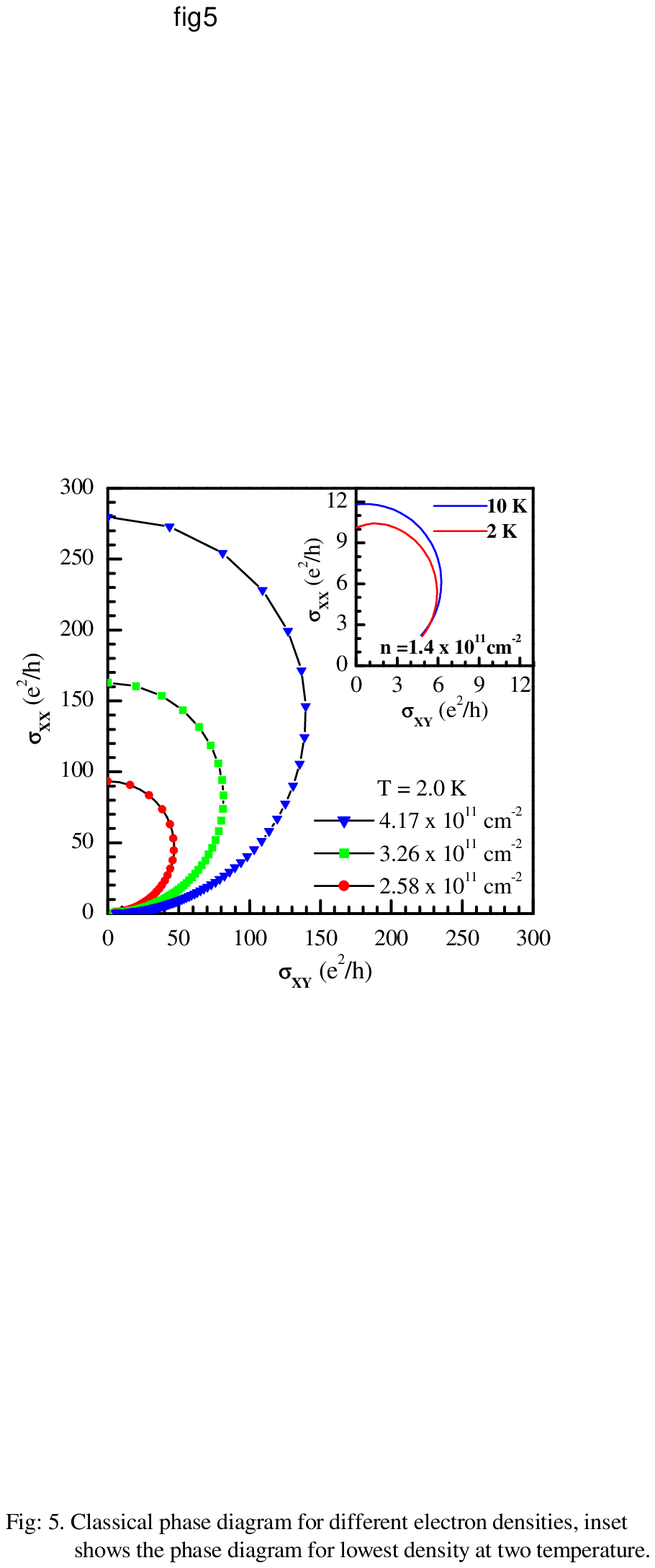}
\end{center}
\caption{\label{f5} Experimental data (symbols) and semiclassical
values (solid lines) in the $\sigma_{xx}$, $\sigma_{xy}$
conductivity plane. The inset shows the different, $T$ dependent
behavior of the experimental data (solid lines) corresponding the
to the lowest electron density.}
\end{figure}

In Fig. 5 we plot the results in units ($e^2 /h$) in the
$\sigma_{xx}$, $\sigma_{xy}$ conductivity plane. We observe that
the data lie on nearly perfect semicircles except those
corresponding to $n=1.4 \times 10^{11} /cm^2$ which are plotted
separately, in the inset of Fig. 5. The semicircles are, in fact,
predicted by the well known results of Drude-Zener-Boltzmann
theory
\begin{equation}\label{scxxy}
\sigma_{xx} =
\frac{\sigma_{0}}{1+(\omega_{c}\tau)^{2}}\hspace{.5in}
\sigma_{xy} =
\frac{\sigma_{0}\omega_{c}\tau}{1+(\omega_{c}\tau)^{2}}.
\end{equation}
with $\sigma_{0} = 1/ \rho_0 = n e^2 {\tau}/m^*$. Eq. (92) implies

\begin{equation}
\left( \sigma_{xx} - \frac{\sigma_0}{2} \right)^2 + \sigma_{xy}^2
=\left( \frac{\sigma_0}{2} \right)^2 ,
\end{equation}
which fits the data in Fig. 5 remarkably well. Several comments
are in order.

\noindent{(i)} Plots like Fig. 5 clearly demonstrate the fact that
the semiclassical Drude-Zener-Boltzmann theory is in many ways a
surprisingly successful theory over the entire range of
experimental $B$. Many interesting and fundamental aspects of the
electron transport problem are being suppressed, however. For
example, the deviations from the semiclassical behavior, like the
{\em nonlinear} effects mentioned in the discussion of Fig. 4, can
no longer be observed when a data set of the same sample taken
over a wide B range is plotted as in Fig. 5.

\noindent{(ii)} Similar statements apply to the fundamental
aspects of the {\em quantum Hall regime}. These are all invisible
in experimental plots of $\sigma_{xx}$ and $\sigma_{xy}$ where the
scale is set by the typical values of the semiclassical theory. As
is well known, the quantum Hall regime generally appears at strong
values of $B$ only. This regime collapses, in Fig. 5, into an
extremely small region around the origin, $\sigma_{xx} \lesssim
1$\cite{kivelson:prb92}.

\noindent{(iii)} All this merely illustrates the well known fact
that quantum Hall regime is experimentally well separated from the
semiclassical regime where the physics is mainly dominated by {\em
weak localization and interaction phenomena} (Section 3.2.5 below). At
strong values of $B$ the problem is in many ways very different
and the role played by the macroscopic inhomogeneities will have
to be considered separately (Section 3.3).

\vskip 3mm
\subsubsection{Semiclassical versus weak quantum interference phenomena}
Next we focus on the low density data in Fig. 5 that have been
plotted separately, in the inset. These data, unlike those taken
at a higher values of the electron density, indicate a significant
departure from the semiclassical theory of electron transport
({\em semicircle law}) as $T$ is lowered. This behavior, with
varying $T$, is naturally explained within the aforementioned
theory of {\em weak localization and interaction} effects. For
this purpose we recall that the quantum theory can in general be
written as in Eq. \ref{weakloc}, i.e.

\begin{equation}\label{spin}
\sigma_{xx} (T) = \sigma_{xx}^c + \beta \ln (T/T_f )
\end{equation}
where $\beta$ and $T_f$ are positive constants that depend on $B$,
electron spin etc. \cite{Finkel}. We have introduced a superscript
'$c$' in order to make an explicit distinction between the {\em
semiclassical} theory on the one hand, and the {\em quantum}
theory of conductances on the other. It is well understood by now
that the theory of {\em weak quantum interference} phenomena
predicts, at the same time, that the Hall component $\sigma_{xy}$
remains unchanged. We can write
\begin{equation}\label{scxy}
\sigma_{xy} (T) = \sigma_{xy} = \sigma_{xy}^c .
\end{equation}
For comparison we have plotted the data for $\sigma_{xx}$ and
$\sigma_{xy}$ for two different $T$ and with varying values of
$\omega_c \tau$ as in Fig. 6. We have made use of the relation

\begin{equation}
\omega_c \tau =B/B_{max}
\end{equation}
where $B_{max}$ is defined as the value of $B$ where the data for
$\sigma_{xy}$ pass through a maximum.

\begin{figure}[h]
\begin{center}
\includegraphics[clip, width=8.0 cm]{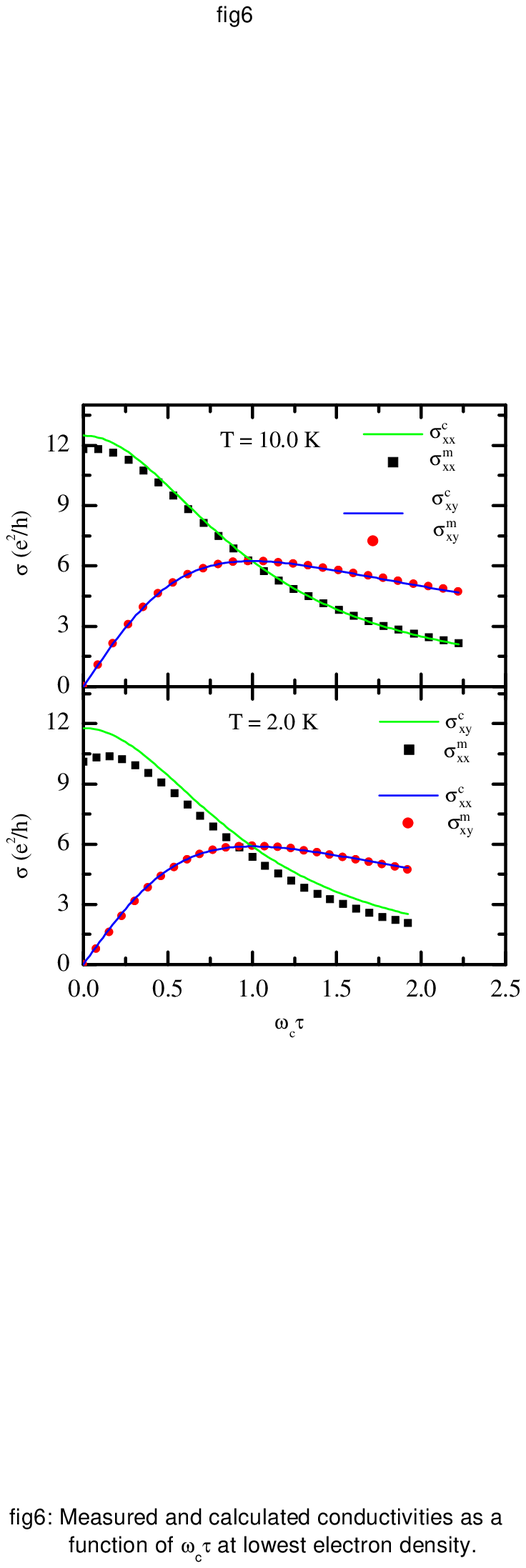}
\end{center}
\caption{\label{f6} Experimental data (symbols) for the
conductivity components $\sigma_{xx}$, $\sigma_{xy}$ for the
sample with lowest density, with varying {$\omega_{c}\tau$} and at
two different values of $T$. The solid lines indicate the
associated semiclassical values (see text). }
\end{figure}

To check the validity of Eq.\ref{scxy} we have fitted the data for
$\sigma_{xy}$ with the semiclassical expression, Eq \ref{scxxy}. The best
fit (solid line in Fig. 6) is obtained by fixing the value of the
parameter $\sigma_0$ at half the maximum value of $\sigma_{xy}$
and the results are listed in Table 1. Once the value of
$\sigma_0$ is known, the semiclassical expression for
$\sigma_{xx}$ is also fixed. These results have been plotted in
Fig. 6 (solid line) as well. On the basis of Fig.6 we can make the
following qualitative statements.

\vskip 2mm

\noindent{(i)} Both the statements of Eqs \ref{spin} and \ref{scxy} are
completely consistent with the experiment. Whereas the data for
$\sigma_{xy}$ follow the semiclassical predictions quite well, the
data for $\sigma_{xx}$ show, as mentioned earlier, a significant
departure from the semiclassical line as $T$ is lowered from $10K$
to $2K$.

\vskip 2mm

\noindent{(ii)} Notice, however, that this departure is
strongest at $\omega_c \tau \approx 0$ and weakest in the regime
$0.5 \lesssim \omega_c \tau \lesssim 2.0$. This is a well known
result of {\em weak quantum interference} theory and indicates
that at $B=0$ and $B\neq 0$ the problem generally belongs to a
different universality class, i.e. the coefficient $\beta$ in Eq.
\ref{spin} is different in each case.

\vskip 2mm

\noindent{(iii)} Since the quantum corrections to $\sigma_{xx}^c$
are relatively small, typically O($e^2 /h$) in the range of
experimental $T$, it becomes immediately obvious why similar
corrections are not observed in the other data in Fig. 5, taken
from the sample at higher values of the electron densities.

\vskip 2mm

\noindent{(iv)} Finally, it is important to emphasize that a
quantitative assessment of scaling phenomena like Eq.\ref{spin} is
complicated, mainly because of the fact that the semiclassical
parameters of the system at low densities depend on $T$ as well.
In particular, the parameter $\sigma_0$ (Table 1) and, hence,
$\sigma_{xx}^c$ and $\sigma_{xy}^c$ in Eqs \ref{spin} and \ref{scxy} vary by
$7\%$ in the range $2K < T < 10K$. Notice that this dependence on
$T$ accounts for roughly $30\%$ of the $T$ dependence as observed
in the $\sigma_{xx}$ data at $\omega_{c}\tau = 0$ (Fig. 6) .

\vskip 2mm

A $T$ dependence of the semi classical parameters has been observed
previously, on similar systems at low densities\cite{Minkov:prb01}.
It indicates that the relaxation time $\tau$ and, hence, the ionized
impurity distribution varies with varying $T$. Physically this happens
because of redistribution of electrons in the doped layer which reduces
the remote ion impurity scattering at higher temperatures.

\vskip 2mm

In summary we can say that transport studies of the semiclassical
regime provide invaluable physical information on the electron gas
that cannot be obtained otherwise. It is quite obvious, for
example, that the type of $T$ dependence in the sample
characteristics as discussed in the present Section can have
dramatic and misleading consequences for the experiments conducted
at strong $B$. In particular, it can be one of the physical
mechanisms that generally prevents one from observing the {\em
universal features} of {\em quantum criticality} of the quantum
Hall plateau transitions, notably the algebraic dependence
$T^\kappa$.

\begin{table}
\begin{center}
\caption{Numerical value of parameters entering into the
semiclassical theory (see text).}
\begin{tabular}{|c|c|c|c|c|} \hline
Temp & {$\bf B_{max}$} & {$\tau$} & {$\sigma_{0}$} & n \\
(K) & (Tesla) & (ps) & {$(\frac{e^{2}}{h})$} &
{$(\times10^{11}cm^{-2})$} \\
\hline 2.0 & 0.52 {$\pm$}0.04 & 0.39 {$\pm$}0.03 & 11.78 {$\pm$}0.04 & 1.48 {$\pm$}0.12 
\\
\hline 10.0 & 0.45 {$\pm$}0.04 & 0.45 {$\pm$}0.04 & 12.48 {$\pm$}0.04 & 1.36 {$\pm$}0.12 
\\
\hline 2.0 & 0.11 {$\pm$}0.01 & 1.20 & 62.80 & 2.58 \\
\hline 2.0 & 0.09 {$\pm$}0.01 & 2.46 & 162.47 & 3.26 \\
\hline 2.0 & 0.07 {$\pm$}0.01 & 3.32 & 279.76 & 4.17 \\
\hline
\end{tabular}
\end{center}
\end{table}

\subsubsection{Numerical values of semiclassical
parameters}

For completeness we next present numerical estimates for the
various parameters that enter into the semiclassical theory.
Specifically we list, in Table 1, the values of $B_{max}$, $\tau$,
$\sigma_0$ and $n$ for each case separately.

First, as far as the data taken at higher densities ($n=2.58,~3.26
,~4.17 \times~ 10^{11} cm^{-2}$) are concerned, these parameters
can be extracted in a standard manner simply because the quantum
corrections are negligible in this case. Specifically, we have
extracted the values for $\sigma_0$ and $n$ directly from the
experimental resistance data by using Eq.\ref{r0h} as well as the
relations $\rho_0 = 1/\sigma_0$ and $\rho_H = B/ne$. Once the
$\sigma_0$ and $n$ are fixed we can accurately compute the
value of $\tau=\sigma_0 m^*/ne^2$ as well. In the computation of $\tau$ the effective
mass $m^*$ of system is considered $0.036m_{e}$. The results do not vary
significantly in the range $2K < T < 10K$. Along with the measured
$B_{max}$ they are listed in Table 1.

Next, for the data corresponding to the lowest value of the
electron density ($n= 1.48,~ 1.36 ~\times~ 10^{11} cm^{-2}$) one
has to follow a slightly different extraction procedure. According
to the discussion following Eq.\ref{scxy} one can use the $\sigma_{xy}$
data and obtain an accurate value for $\sigma_0$ and an estimate
for $B_{max}$ which is generally much less accurate, however.
Given the values of $\sigma_0$ and $B_{max}$ one can next compute
the electron density $n=\sigma_{0} B_{max}/e$ as well as the
relaxation time $\tau= m^*/e B_{max}$. The accuracy of the
computation is mainly limited by experimental uncertainties in
$B_{max}$. The results obtained for $T=2K$ and $10K$ are listed in
Table 1. Notice that the difference in $n$ at $2K$ and $10K$ is
insignificant in view of the large error bars. On the other hand,
we do attribute the $T$ dependence in $\sigma_0$ to the fact that
$\tau$ is $T$ dependent.

\begin{figure}[h]
\begin{center}
\includegraphics[clip, width=8.0 cm]{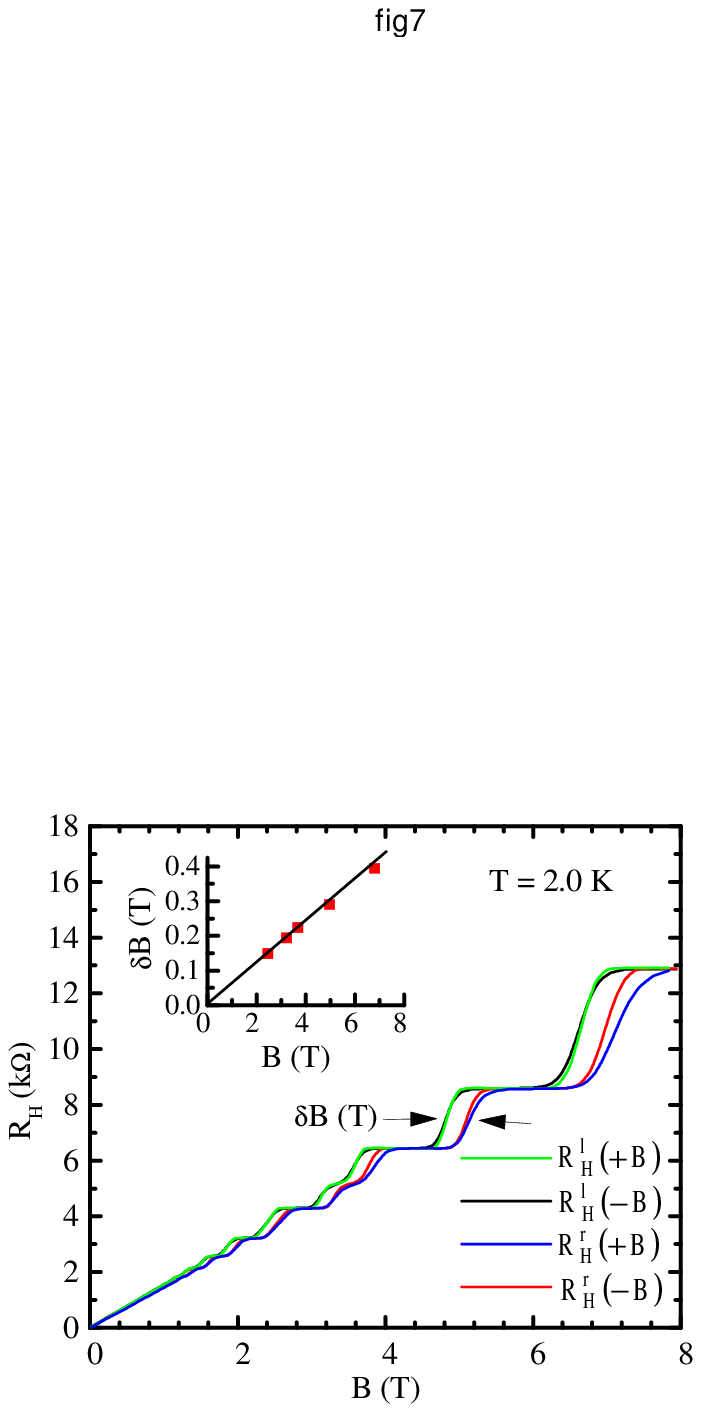}
\end{center}
\caption{\label{f7} High field Hall resistance, inset shows the
difference of critical magnetic field for PP transitions }
\end{figure}

\subsection{Quantum Hall regime}
\subsubsection{Aspects of symmetry}
We next consider the transport properties in the regime of high
$B$ where the quantum features dominate for average electron density
$4.17 \times~ 10^{11} cm^{-2}$. Fig. 7 shows the
magnitude of Hall resistances $R_H^l$ and $R_H^r$ as measured at
the contacts  $AB$ and $CD$ of the Hall bar with varying $B$ up to 8
Tesla and for both directions of the field. Fig. 8 shows the
corresponding data of the longitudinal resistances $R_0^t$ and
$R_0^b$ as measured at the contacts $AC$ and $BD$ respectively.

\begin{figure}[h]
\begin{center}
\includegraphics[clip, width=8.0 cm]{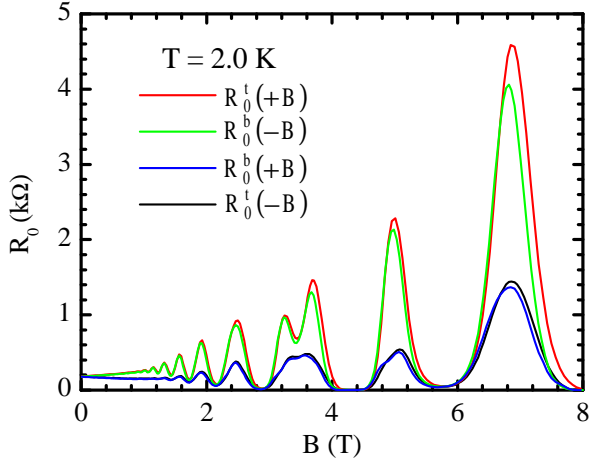}
\end{center}
\caption{\label{f8} High field longitudinal magnetoresistance}
\end{figure}

As a first investigation of inhomogeneity effects we study the
symmetry of the data under a change in the {\em polarity} of the
magnetic field $B$. Following Eqs (15-17) the two sets of data for
$R_0^t$ and $R_0^b$ should be related according to
\begin{equation}
R_{0}^{t}(+B)=R_{0}^{b}(-B),~~ R_{0}^{t}(-B)=R_{0}^{b}(+B) .
\end{equation}
Similarly, the two sets of data for $R_h^l$ and $R_H^r$ should be
related as follows

\begin{equation}
R_H^{l}(+B)= - R_{H}^{l}(-B),~~ R_{H}^{r}(+B)= - R_{H}^{r}(-B) .
\end{equation}
From Figs 7 and 8 we observe that the statements of symmetry are
all in qualitative agreement with the experiment at high $B$,
indicating that the gradients in the electron density still remain
the principle reason for having imperfect data.

\subsubsection{Numerical estimate for density gradients}
The shift in the "critical" $B$ values for the PP transition as
taken from the left contact ($AB$) and the right contact ($CD$) is
directly related to the filling fraction $\nu = nh/eB$ of the
Landau level system. Since the electron density $n$ varies along
the channel of the Hall bar, the PP transition at the low-density
side (AB) occurs at a relatively lower value for $B$ whereas at
the high-density side (CD) the transition takes place at a
relatively higher $B$. Notice that this is precisely the statement
made by Eq. (54) which says that for fixed $T$ and varying $B$ the
measured Hall resistances can be written as follows

\begin{equation}
R_H^l (B) = \rho_H (n_l ,B), ~~~~R_H^r (B) = \rho_H (n_r ,B) .
\end{equation}
Here, $n_l$ and $n_r$ are the different but fixed electron
densities at the left contacts ($AB$) and right contacts ($CD$) of
the Hall bar respectively. Equating $R_H^l$ and $R_H^r$ amounts to
taking different values $B_l$ and $B_r$ for the magnetic field $B$
at the left and the right hand side of the Hall bar such that the
filling fraction is the same. Hence

\begin{equation}
\nu = \frac{n_l h}{e B_l} = \frac{n_r h}{e B_r}
\end{equation}
from which it follows immediately that

\begin{equation}
\frac{\delta n}{n}  =  2 \frac{n_r - n_l}{n_r + n_l} = 2 \frac{B_r
- B_l}{B_r +B_l} = \frac{ \delta B}{B}
\end{equation}
irrespective of whether the Landau levels are spin polarized or
not.

In the inset of Fig. 7 we have taken $\delta B$ as the difference
in the value of $B$ where the data for $R_H^l$ and $R_H^r$ pass
through the "center" of the plateau transition. We observe that
the $\delta B$ as taken from the different plateau transitions
varies linearly with $B$. The slope of the line gives an estimate
for the relative difference $\delta n/n$ in the electron densities
at the left and right contacts of the Hall bar. More specifically
we have ${\delta}B/B = {\delta}n/n = 6{\%}$ which agrees
remarkably well with the aforementioned result ($ 6.2 \%$) as obtained
from the low $B$ data.

\subsubsection{Beyond linear approximation}
We have seen so far that the transport data at high $B$ do indeed
display the features of symmetry as predicted by Eqs (15-17). At
the same time one is also able to obtain reliable numerical
estimates for the density gradients from the experimental data for
both the quantum Hall regime and the semiclassical regime. This,
however, does not take away from the fact that additional features
are clearly present in the data that are caused by different
aspects of inhomogeneity. In this Section we shall point out that
a proper analysis of the experiment on the PP transitions is
actually way beyond the limitations of the {\em linear
approximation}. This statement becomes most obvious by looking at
the quantum Hall plateau transitions in the $R_H^l$ and $R_H^r$
data (Fig. 7). Recall that the plateau transitions as observed in
the $R_H^l$ and $R_H^r$ data are shifted by amount $\delta B$ in $B$
due to density gradients. This $\delta B$, however, is comparable
or even larger than the actual {\em width} $\Delta B$ of the
plateau transition as observed in the same data. Notice that
$\Delta B \propto T^{\kappa}$ is the quantity of actual
interest whereas the ratio $\frac{\delta B}{\Delta B(T)}$ is
identically the same as the ratio $\frac{\delta \nu_x}{\nu_0 (T)}$
introduced in Section 2.4. We therefore conclude that the present
experiment is conducted in a regime of $T$ where one can no longer
expect the data to provide reliable information on the quantum
critical behavior of the electron gas, notably the numerical value
of the critical index $\kappa$. This conclusion is entirely
consistent with the experimental data on $R_0^t$ and $R_0^b$ (Fig.
8). In particular, the measured $R_{0}^{t,b}$ provide a
substantially larger estimate for the width $\Delta B$ of the PP
transitions than what one obtains from the $R_H^{l,r}$ data. This
discrepancy can simply be understood from the fact that the
longitudinal resistance is measured along the length of the Hall
bar where the electron density changes the most. $R_{0}^{t,b}$
therefore represents an {\em average} over a range of electron
densities. This is unlike the quantity $R_H^{l,r}$ which probes
the local electron densities at the left hand side and right hand
side of the Hall bar, $n_l$ and $n_r$ respectively.

These considerations simply illustrate the fact that the resistance
data do not necessarily reflect the intrinsic properties of the
quantum phase transition alone. Macroscopic sample inhomogeneities
may enter the problem in a highly non-linear fashion and, hence,
they may fundamentally complicate the studies of quantum
criticality in the quantum Hall regime.

\begin{figure}[h]
\begin{center}
\includegraphics[clip, width=8.0 cm]{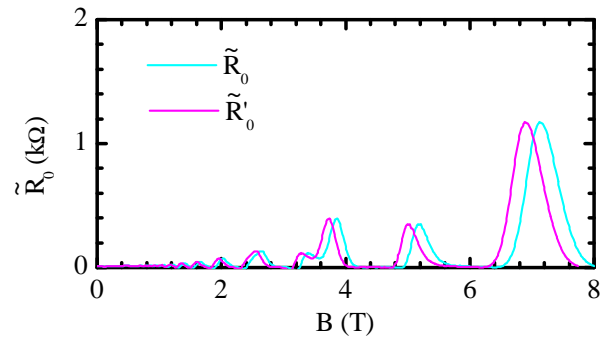}
\end{center}
\caption{Plot of difference resistance $\tilde{R}_0$ extracted
from Hall resistances at the CD contacts (Fig. 1). Also the difference
resistance $\tilde{R'}_0$ with B values rescaled is plotted (see text)}
\end{figure}

\subsubsection{Contact misalignment}
In this Section we address the small deviation from the ideal
reflection symmetry, Eqs. (15-17), that can clearly be seen in the
experimental data on the Hall resistance, $R_H$, in Fig. 7.
Whereas the data for $R_H^l$ nicely display the symmetry under the
change $B \rightarrow -B$, this is not so for the $R_H^r$ data
which have been measured from the {\em right} hand side contacts
($CD$, Fig. 1) of the Hall bar. This asymmetry between the
$R_H^r (B)$ and $R_H^r (-B)$ signals can easily be explained,
however, if one assumes that the {\em right} hand side contacts
$CD$, unlike the $AB$ on the {\em left} hand side, are slightly
{\em misaligned}. More specifically, by assuming that the contacts
$C$ and $D$ are misaligned in the x-direction by a small amount,
say $\Delta << L$, then the difference between the Hall
resistances, $\tilde{R}_0 = |R_H^r (-B)| - |R_H^r (B)|$, should
be proportional to the local (near CD, Fig. 1) longitudinal resistance $R_0$.

To test the idea of contact misalignment we have plotted, in
Fig.9, the quantity $\tilde{R}_0$ with varying $B$. As expected,
the results look very much the same as the experimental plot of
$R_0$, Fig. 8. One can build upon these results in several ways.
For example, one thing to notice is that the {\em peaks} in the
$\tilde{R}_0$ data corresponding to the different PP transitions
are shifted towards higher $B$ values as compared to the $R_0$
data in Fig. 8. This shift in $B$ is a result of the fact that the
$\tilde{R}_0$ actually probe the local longitudinal resistance
of the system at CD contacts where as $R_0$ represents an average
over a range of electron densities. For the purpose of showing this
we have also plotted the results (denoted by $\tilde{R'}_0$ in Fig. 9)
that have been obtained from $\tilde{R}_0$ after a simple re-scaling
of the $B$ axis with a factor $\frac{n_r + n_l}{2n_r} =0.967$. Here,
$n_r$ and $n_l$ are the electron densities at opposite sides of the
Hall bar that we have considered earlier. The peaks in $\tilde{R'}_0$
and those in $R_0$ now occur at approximately the same values for
$B$, indicating that the effect is dominated by the variations in
the electron density that are {\em linear} in the spatial coordinates.
For completeness we report the numerical value of the misalignment
$\Delta$. On the basis of the magneto resistance measurements at
low $B$ we conclude that $\Delta$ is equal to $2.5 \%$ of the length
of the Hall bar, $L$ (Fig. 1).

\section{Summary and conclusions}

We have classified and analyzed the most important effects of
sample inhomogeneity on the transport data taken from low mobility
heterostructures. We have shown, in particular, that the presence
of density gradients manifests itself most clearly in terms of a
{\em reflection symmetry} that generally exists in the
longitudinal and transverse magneto resistance data taken from the
symmetrically placed contacts on the Hall bar.

The expressions obtained under the so-called {\em linear
approximation} allow us to extract the main features of
semiclassical transport and weak quantum interference, even for
samples containing a density gradient of nearly $6$ percent. Going
beyond the linear approximation, we have introduced an exactly
solvable model with exponential density gradients. The {\em
nonlinear inhomogeneity effects} are found to be weak in the
regime of semiclassical transport or weak $B$, but they generally
complicate the experiments on scaling of the PP transitions. Since
the sample used in the present experiments possesses density
gradients which are rather large, it clearly brings out the
futility of attaching meaning to the shapes and $T$ dependence of
the magneto resistance measurements on the PP transitions. While
it is best therefore to avoid the complications of sample
inhomogeneities altogether, the analysis specifies fundamental
bounds on the experimental density gradients which may be
tolerated in the study of quantum criticality of both the PP and
PI transitions. It is further seen that the experimental
restrictions are far more severe for the PP transitions than for
the PI transition taken from the same sample. These restrictions
mainly place a lower limit on the experimental $T$ below which the
quantum critical behavior of the electron gas can no longer be
studied. Along with the density gradients, our present experiments
also reveal the effects of another common aspect of inhomogeneity,
the misalignment of the Hall bar contacts, that generally affects
the shape and $T$ dependence of the quantum transport measurements
at the PP and PI transitions.
\acknowledgments
One of us (B.K.) acknowledges the Kanwal Rekhi
Scholarship of the TIFR Endowment Fund for partial support. Two of
us (B.K. and A.M.M.P.) were funded, in part, by a grant from FOM.

\appendix


\section{Aspects of symmetry}


In this Appendix we point out that the universality of the scaling
functions, Eqs (72-76), is actually a statement of universality
made on the $\beta$ and $\gamma$ functions of the electron gas.
For this purpose we recall that the Coulomb interaction problem
generally involves the renormalization  of three distinct
parameters, namely the dimensionless conductances $\sigma_{xx}$
and $\sigma_{xy}$ as well as the {\em singlet interaction
amplitude} $z$ that is associated with the variable $T$ and/or the
external frequency. The most fundamental quantities of the theory
are defined as follows

\begin{eqnarray}
\frac{d \sigma_{xx}}{d \ln \mu} &= & \beta_{xx}
(\sigma_{xx},\sigma_{xy}) \\
\frac{d \sigma_{xy}}{d \ln \mu} &= & \beta_{xy}
(\sigma_{xx},\sigma_{xy}) \\
\frac{d \ln zT}{d \ln \mu} &= & 2+ \gamma
(\sigma_{xx},\sigma_{xy}).
\end{eqnarray}
Here, $\mu$ denotes an arbitrary momentum scale in the problem and
$\gamma (\sigma_{xx},\sigma_{xy})$ is known as the {\em anomalous
dimension} of the variable $z$ or $T$. The following symmetries
are fundamental features of quantum Hall systems

\noindent{\em $\bullet$ ~Particle-hole symmetry~~}
\begin{eqnarray}
\beta_{xx} (\sigma_{xx},\sigma_{xy}) & = & ~\beta_{xx}
(\sigma_{xx}, 1- \sigma_{xy})\\
\beta_{xy} (\sigma_{xx},\sigma_{xy}) & = & - \beta_{xy}
(\sigma_{xx}, 1- \sigma_{xy})\\
\gamma (\sigma_{xx},\sigma_{xy}) & = & ~~~\gamma (\sigma_{xx}, 1-
\sigma_{xy}) .
\end{eqnarray}

\noindent{\em $\bullet$ ~Periodicity in $\sigma_{xy}$~~}
\begin{eqnarray}
\beta_{xx} (\sigma_{xx},\sigma_{xy}) & = & ~\beta_{xx}
(\sigma_{xx}, \sigma_{xy} +k)\\
\beta_{xy} (\sigma_{xx},\sigma_{xy}) & = & ~ \beta_{xy}
(\sigma_{xx}, \sigma_{xy} +k)\\
\gamma (\sigma_{xx},\sigma_{xy}) & = & ~~~\gamma (\sigma_{xx},
\sigma_{xy} +k) .
\end{eqnarray}
These symmetries give rise to the general statement which says
that the quantum Hall plateaus are described by a series of
equivalent {\em stable} fixed points located at $\sigma_{xx} = 0$
and $\sigma_{xy} = k$. At the same time, the critical
singularities of the plateau transitions are controlled by a
series of equivalent {\em unstable} fixed points located at
precisely half-integer values of the Hall conductance.

\vskip 3mm

\section{Experimental $\beta$ functions}

To establish the contact with the experiment we introduce the
following quantities

\begin{eqnarray}
\frac{d \sigma_{xx}}{d \ln zT} & = & \frac{\beta_{xx}
(\sigma_{xx},\sigma_{xy})}{2+ \gamma (\sigma_{xx},\sigma_{xy})} =
\tilde{\beta}_{xx}
(\sigma_{xx},\sigma_{xy}) \\
\frac{d \sigma_{xy}}{d \ln zT} & = & \frac{\beta_{xy}
(\sigma_{xx},\sigma_{xy})}{2+ \gamma (\sigma_{xx},\sigma_{xy})} =
\tilde{\beta}_{xy} (\sigma_{xx},\sigma_{xy}) .
\\ \nonumber
\end{eqnarray}
Notice that the experimental $\tilde{\beta}_{xx}$ and
$\tilde{\beta}_{xy}$ functions display the same symmetries as the
original ones, $\beta_{xx}$ and $\beta_{xy}$.

Next, on the basis of Eqs (72) and (73), which are defined on the
interval $0\leqq \sigma_{xy} \leqq 1$, one can obtain the explicit
expressions for $\tilde{\beta}_{xx}$ and $\tilde{\beta}_{xy}$ as
follows. First we solve for the quantities $X$ and $\eta$

\begin{equation}
X = \Phi_{odd} (\sigma_{xx},\sigma_{xy}) ;~~~ \eta = \Phi_{even}
(\sigma_{xx},\sigma_{xy}) ,
\end{equation}
where

\begin{eqnarray}
\Phi_{even} (\sigma_{xx},\sigma_{xy}) & = & \frac{ \frac{1}{2} -
\sqrt{(\sigma_{xy} -\frac{1}{2}
)^2 + \sigma_{xx}^2 }}{\sigma_{xx} } \\
\Phi_{odd} (\sigma_{xx},\sigma_{xy}) & = & \ln \left(
\frac{\sigma_{xy} - \sigma_{xx} \Phi_{even} }{\sigma_{xx} }
\right) .
\end{eqnarray}
Notice that

\begin{eqnarray}
\Phi_{even} (\sigma_{xx},\sigma_{xy}) & = & ~~\Phi_{even} (\sigma_{xx},1-\sigma_{xy}) \\
\Phi_{odd} (\sigma_{xx},\sigma_{xy})& = & -\Phi_{odd}
(\sigma_{xx},1-\sigma_{xy}).
\end{eqnarray}
Apparently we have

\begin{eqnarray}
\frac{d \Phi_{odd}}{d \ln T} & = & -\kappa \Phi_{odd} \\
\frac{d \Phi_{even}}{d \ln T} & = & y_\sigma \Phi_{even} ,
\end{eqnarray}
such that $\Phi_{odd}$ and $\Phi_{even}$ can be identified as the
Wegner scaling fields in the problem.

The experimental $\tilde{\beta}$ functions in the interval $0\leqq
\sigma_{xy} \leqq 1$ are now obtained as follows
\begin{eqnarray}
 && \tilde{\beta}_{xx} (\sigma_{xx},\sigma_{xy}) = 
+2 \sigma_{xx} \times \nonumber \\ 
&&\left\{ \kappa (\sigma_{xy} - \frac{1}{2} ) \Phi_{odd}  
- y_{\sigma}\sigma_{xx} \Phi_{even} \right\} \\
&& \tilde{\beta}_{xy} (\sigma_{xx},\sigma_{xy}) = 
-2 \sigma_{xx} \times \nonumber \\ 
&&\left\{ \kappa (\sigma_{xx} + \frac{1}{2}\Phi_{even} ) 
\Phi_{odd} + y_{\sigma} ( \sigma_{xy} -
\frac{1}{2} ) \Phi_{even} \right\} \qquad.
\end{eqnarray}
First, it is readily established that the result satisfies {\em
particle-hole symmetry} as it should be

\begin{eqnarray}
\tilde{\beta}_{xx} (\sigma_{xx},\sigma_{xy}) & = &
~\tilde{\beta}_{xx}
(\sigma_{xx}, 1- \sigma_{xy})\\
\tilde{\beta}_{xy} (\sigma_{xx},\sigma_{xy}) & = & -
\tilde{\beta}_{xy} (\sigma_{xx}, 1- \sigma_{xy}).
\end{eqnarray}
Secondly, we make use of the statement of {\em periodicity in
$\sigma_{xy}$}, Eqs (77-79), and extend the expressions for
$\tilde{\beta}$ to include the entire range of $\sigma_{xy}$, i.e.
\begin{eqnarray}
\tilde{\beta}_{xx} (\sigma_{xx},\sigma_{xy}) & \rightarrow &
~\tilde{\beta}_{xx}
(\sigma_{xx}, \sigma_{xy} +k)\\
\tilde{\beta}_{xy} (\sigma_{xx},\sigma_{xy}) & \rightarrow &
\tilde{\beta}_{xy} (\sigma_{xx}, \sigma_{xy} +k) .
\end{eqnarray}
As a final step we can next employ the analysis of this Section in
a backwards manner and show that the solutions to the differential
equations, Eqs (B.1) and (B.2), are indeed given by Eq. (77) with
the parameters $T_0^{(k)}$ and $T_1^{(k)}$ being left undetermined.


\end{document}